%% file: Main.tex
\documentclass[conference]{IEEEtran} 
\IEEEoverridecommandlockouts
\usepackage{graphicx}
\usepackage{subfig}
\usepackage{amsmath,amsthm,amsfonts,amssymb}
\usepackage{cite}
\usepackage{bm}
\usepackage{bbm}
\usepackage{url}
\usepackage{array}
\usepackage{color}
\usepackage{multirow}
\usepackage{booktabs}
\usepackage{xcolor}
\usepackage{enumerate}
\usepackage{epstopdf}
\usepackage{tabularx}
\usepackage{algorithm}
\usepackage[english]{babel}
\usepackage{algpseudocode}
\usepackage{listings}
\usepackage{tcolorbox}
\usepackage{float}
\tcbset{
  sharp corners,
  boxrule=0.5pt,
  top=0mm,bottom=0pt,left=0pt,right=0pt,
  boxsep=0.5pt,
}
\tcbuselibrary{skins, breakable}

\definecolor{vgreen}{RGB}{104,180,104}
\definecolor{vblue}{RGB}{49,49,255}
\definecolor{vorange}{RGB}{255,143,102}

\lstdefinestyle{verilog-style}
{
    language=Verilog,
    basicstyle=\tiny\ttfamily,
    keywordstyle=\color{vblue},
    identifierstyle=\color{black},
    commentstyle=\color{vgreen},
    numbers=left,
    numberstyle=\tiny\color{black},
    numbersep=10pt,
    tabsize=2,
    moredelim=*[s][\colorIndex]{[}{]},
    literate=*{:}{:}1
}
\lstdefinestyle{prompt-style}
{
    language=Verilog,
    basicstyle=\tiny\ttfamily,
    keywordstyle=\color{black},
    identifierstyle=\color{black},
    commentstyle=\color{black},
    numbers=left,
    numberstyle=\tiny\color{black},
    numbersep=10pt,
    tabsize=2,
    moredelim=*[s][\colorIndex]{[}{]},
    literate=*{:}{:}1
}

\makeatletter
\newcommand*\@lbracket{[}
\newcommand*\@rbracket{]}
\newcommand*\@colon{:}
\newcommand*\colorIndex{%
    \edef\@temp{\the\lst@token}%
    \ifx\@temp\@lbracket \color{black}%
    \else\ifx\@temp\@rbracket \color{black}%
    \else\ifx\@temp\@colon \color{black}%
    \else \color{vorange}%
    \fi\fi\fi
}
\makeatother

\usepackage{trace}

\begin{document}
\title{\makebox[\linewidth]{\parbox{\dimexpr\textwidth+0.975cm\relax}{\centering The Power of Large Language Models for Wireless Communication System Development: A Case Study on FPGA Platform}}}

\author{
    \IEEEauthorblockN{Yuyang Du, Hongyu Deng, Soung Chang Liew$^*$, Yulin Shao, Kexin Chen, He Chen} \vspace{-6em}
    \IEEEauthorblockA{\thanks{Y. Du, H. Deng, S. C. Liew, K. Chen, and H. Chen are with the Chinese University of Hong Kong, Hong Kong. Y. Shao is with the University of Macau, Macau.}} 
    \IEEEauthorblockA{\thanks{*{Corresponding author: S. C. Liew (soung@ie.cuhk.edu.hk)}.}}
}
\maketitle

\renewcommand{\arraystretch}{1.25}

\begin{abstract}
Large language models (LLMs) have garnered significant attention across various research disciplines, including the wireless communication community. There have been several heated discussions on the intersection of LLMs and wireless networking technologies. While recent studies have demonstrated the ability of LLMs to generate hardware description language (HDL) code for simple computation tasks, developing wireless networking prototypes/products via HDL poses far greater challenges because of the more complex computation tasks involved. In this paper, we aim to address this challenge by investigating the role of LLMs in FPGA-based hardware development for advanced signal-processing algorithms in wireless communication and networking. We begin by exploring LLM-assisted code refactoring, reuse, and validation, using an open-source software-defined radio (SDR) project as a case study. Through the case study, we find that an LLM assistant can potentially yield substantial productivity gains for researchers and developers. We then examine the feasibility of using LLMs to generate HDL code for advanced wireless signal processing, using the Fast Fourier Transform (FFT) algorithm as an example. This task presents two unique challenges: the scheduling of subtasks within the overall task and the multi-step thinking required to solve certain arithmetic problem within the task. To address these challenges, we employ in-context learning (ICL) and Chain-of-Thought (CoT) prompting techniques, culminating in the successful generation of a 64-point Verilog FFT module. Our results demonstrate the potential of LLMs for generalization and imitation, affirming their usefulness in writing HDL code for wireless networking systems. Overall, this work contributes to understanding the role of LLMs in wireless networks and motivates further exploration of their capabilities.
\end{abstract}


\section{Introduction}\label{sec-I}
\input{./Source/Section_I.tex}

\section{Background and Related Works}\label{sec-II}
\input{./Source/Section_II.tex}

\section{Using LLMs to Assist FPGA Development}\label{sec-III}
\input{./Source/Section_III.tex}

\section{Challenges in implementing Complex Signal-processing Algorithm in Verilog}\label{sec-IV}
\input{./Source/Section_IV.tex}

\section{Solving implementation challenges via ICL and CoT Prompting}\label{sec-V}
\input{./Source/Section_V.tex}

\section{Experimental Investigation of the Productivity Gain}\label{sec-new}
\input{./Source/Section_New_Benchmarking.tex}

\section{Conclusion}\label{sec-VI}
\input{./Source/Section_VI.tex}

\bibliographystyle{IEEEtran}
\bibliography{./Reference/ref}

\appendices
\renewcommand\thefigure{\thesection\arabic{figure}}
\renewcommand\thetable{\thesection\arabic{table}}

\newpage
\section{Prompts and Source Codes}\label{sec-App1}
\input{./Source/Appendix_A1.tex}
\input{./Source/Appendix_A2.tex}
\input{./Source/Appendix_A3.tex}

\input{./Source/Appendix_A4.tex}
\input{./Source/Appendix_A5.tex}
\input{./Source/Appendix_A6.tex}

\newpage
\section{Supplementary Experiments}\label{sec-App2}
\input{./Source/Appendix_B1.tex}
\input{./Source/Appendix_B2.tex}
\input{./Source/Appendix_B3.tex}
\end{document}

%% file: Source/Section_I.tex
The emergence of large language models (LLMs) has garnered significant attention within the research community. Scholars from diverse scientific disciplines are intrigued by LLMs due to their potential to transcend the realm of natural language processing (NLP). Researchers from the wireless communication and networking field have also taken notice of this trend, leading to a series of insightful discussions in the community \cite{ref001,ref002,ref003}.

Recent studies indicated that LLMs possess the ability to generate hardware description language (HDL), such as Verilog, thereby painting a promising picture of LLMs aiding researchers and hardware engineers in the foreseeable future. The advancement of LLM-written HDL is particularly encouraging for the wireless networking community, given that many experimental and systematic studies in the field prototype their system in field programmable gate arrays (FPGA) platforms that are programmed using HDL \cite{refB1,refB2,refB3,refB4,refB5}. With the assistance of LLMs, researchers are expected to achieve a more rapid system development on FPGA with less manpower consumed.

However, FPGA projects in wireless communication involve complex signal processing algorithms and are therefore difficult to develop. Prior works on LLM-assisted Verilog programming \cite{refC1,refC2,refC3,refC4,refC5,refC6,refC7,refC8} only reported simple hardware logics, such as those for shift register and multi-function calculators with addition/subtraction/multiplication/division (see Section \ref{sec-II}-C for a detailed analysis). These operations are only small atomic computational tasks, whereas the algorithms for signal processing blocks in wireless networking are far more complex. Knowing only the Verilog code for these atomic operations does not guarantee successful Verilog code for the higher-level signal processing blocks. This raises a critical question: How could LLMs assist in the development of FPGA projects involving intricate wireless signal-processing algorithms? This question remains open and calls for further investigation. This paper presents our endeavors in addressing this compelling issue, contributing to the growing discourse on the potential intersection between LLMs and wireless networking technologies.

This paper contributes to two pertinent topics. \textit{In our first contribution}, we investigate the potential of LLMs as a valuable tool for wireless researchers engaged in FPGA-based prototype/product development. To this end, we examine an open-source, FPGA-based SDR project \cite{refB1} as a case study. We thoroughly analyze the project's code and conduct comprehensive experiments to explore the extent to which an LLM can assist in implementing a wireless system. Our examination identifies three pivotal uses of LLMs: code refactoring, code reuse, and code validation. These uses of LLM, while seemingly mundane, are indispensable in hardware development and point towards the capacity of LLMs to substantially amplify researchers' productivity and expedite their research and development process.

\textit{In our second contribution}, we delve into the possibility of employing LLMs to generate sophisticated HDL code for advanced signal processing algorithms required in wireless communication. To illustrate this, we focus on the Fast Fourier Transform (FFT) algorithm as an example. We emphasize that utilizing LLMs to implement FFT in HDL presents significantly more formidable challenges compared to generating code in commonly used languages such as C, Python, or MATLAB (see Section \ref{sec-IV}-A for a detailed discussion).

The first challenge we encountered in using LLMs to generate HDL code for FFT is the scheduling of subtasks inside the FFT module. In general, a complex computation task like FFT can be broken down into a connection of smaller subtasks \cite{ref004}. In high-level languages like C, compilers and operating systems (OSs) can schedule the processing of subtasks to hardware processors in a computer so that programmers do not need to worry about the scheduling issue. However, HDL programming requires designers to interface directly with the hardware without the guidance provided by compilers or OSs. This necessitates a meticulous consideration of parallelism and the precedence relationship between subtasks in HDL programming, a nuance that we have observed to be lacking in current LLMs like ChatGPT. Consequently, using similar prompts as in previous works \cite{refC1,refC2,refC3,refC4,refC5,refC6,refC7,refC8} cannot yield a workable FFT for more than four points. To address the challenge, our prompt design applies in-context learning (ICL). This few-shot learning technique enables an LLM to rapidly learn from the additional examples we give about the parallelism/precedence inside a small-scale FFT (say four-point or eight-point), on which it has not been previously trained.

The second challenge is the limited multi-step thinking ability of LLM. Some recent works \cite{ref005,ref006,ref007,ref017} have reported that LLMs do not perform well when given a complex multi-step task, as they cannot decompose the problem as a human engineer can. In an FFT module, however, there are some complex processes that need to be decomposed into multiple steps, say calculating the twiddle factors and expressing them in the form of signed binary numbers. To augment the multi-step problem-solving ability of LLMs with that of human engineers, we exploit Chain-of-Thought (CoT) prompting. This method teaches LLMs how to approach complex multi-step problems in a way that mimics human thinking, enabling them to handle more complex tasks with greater accuracy and efficiency.

By incorporating the latest research outcomes in NLP (i.e., ICL and CoT prompting) into the FPGA implementation of complex wireless communication algorithms, we achieved a remarkable milestone in this paper: the successful generation of a 64-point Verilog FFT module using LLM. To the best of our knowledge, this is the first LLM-written complex HDL module ever reported in the field. More importantly, our explorations provide valuable insights into the understanding of LLMs:
\begin{enumerate}
    \item LLMs demonstrate remarkable generalization abilities. They can realize sophisticated iterative wireless communication algorithms in HDL, provided that all ambiguities are effectively addressed during the instructional phase.
    \item LLMs exhibit a strong ability to imitate. Once taught the problem-solving approach of a human, they can comprehend complex calculations.
\end{enumerate}
These insights highlight the potential of LLMs and pave the way for leveraging LLMs to write HDL code for wireless communication building blocks.

Additionally, we conducted benchmark experiments involving a wider pool of participants to quantify the productivity gains provided by LLMs in wireless network prototyping. We constructed a set of HDL programming tasks representative of typical challenges. These tasks were distributed to two groups of volunteers: one composed of undergraduates with related coursework experience, and another consisting of senior postgraduate students from an ASIC\&FPGA related lab. For comparison, two equivalently diverse groups (one undergraduate and one postgraduate) were assigned the same tasks, but they could finish these tasks with LLM's assistance. We evaluate each group based on code quality and completion time. The results indicate that the aid of LLMs notably reduces coding time even for experienced engineers and enhances their code quality.

%% file: Source/Section_II.tex
\subsection{Large Language Models (LLMs)}
LLMs leverage the transformer architecture \cite{refA1}. Early research like BERT \cite{refA2} and GPT-2 \cite{refA3} paved the way for today's boom. But it was the advent of GPT-3 \cite{refA4} and its successors that brought the public's attention to the potential of these models. Today, the landscape is diverse, featuring numerous LLMs, with an array of options for both general and task-specific applications.

Despite their variations, all LLMs share core characteristics. First, they all serve as ``scalable sequence prediction models" \cite{refA5}, meaning that they generate the ``most probable" continuation of an input prompt. Second, LLMs operate on ``tokens", which are commonplace character sequences specified through byte pair encoding. This method allows efficient data handling within the constraint of LLMs' fixed context size. By operating over tokens instead of characters, LLMs can process more text. For instance, in OpenAI's models, each token corresponds to roughly 4 characters, and the context windows can accommodate up to 8,000 tokens.

\subsection{FPGA-based SDR Development for Wireless Networking}
Software-defined radios (SDRs) provide a flexible and cost-effective solution to adapt equipment to the fast-evolving wireless communication standards and serve various research purposes. Instead of hardware-centric traditional radios, where different hardware is required to process different signals, SDRs allow the functionalities of a radio system to be defined and altered through software, making it possible to support multiple standards and applications with a single platform.

Thanks to its parallel processing capabilities, reprogrammability, and high performance, FPGA plays a pivotal role in SDR development. It provides an efficient platform for implementing complex, computation-intensive signal processing algorithms essential in SDRs. The application of FPGA-based SDRs applications span across various domains, from cellular networks, WiFi, and satellite communications to specialized underwater communication systems.

FPGA programming involves using an HDL to describe digital circuitry, ranging from simple combinational circuits to intricate sequential circuits and more complex systems-on-chip (SoCs). There are two standard HDLs for FPGA: VHDL and Verilog, with this paper using Verilog as an example.

One of the most widely recognized open-source FPGA-based SDR projects in recent years is OpenWiFi \cite{refB1}, which aims to provide a fully software-defined, reprogrammable WiFi networking solution via Verilog programming, and it has attracted much research interest. OpenWiFi runs on a high-performance Xilinx FPGA board, which provides ample hardware resources to implement the IEEE 802.11 standards.

\subsection{LLM for Hardware Design}
Recent research has shown a growing interest in harnessing the capabilities of LLMs to assist researchers and engineers in hardware design. In \cite{refC1}, the authors employed GitHub Copilot to scrutinize the incidence rates of six types of Verilog bugs. Following this, \cite{refC2} and \cite{refC3} investigated the potential for automated bug repairs with LLMs' assistance. This trend is not limited to academia, as industry players such as RapidSilicon are promoting their upcoming LLM-assisted tool for hardware design, called RapidGPT \cite{refC4}.

Beyond assisting researchers and engineers, recent studies also show interest in replacing human HDL programmers with LLMs. Initial efforts in this direction were documented in \cite{refC5}, where a fine-tuned GPT-2 model was trained with synthetically generated Verilog snippets. However, the limited generalization ability to unfamiliar tasks was a notable shortcoming of \cite{refC5}. Subsequent research \cite{refC6} expanded on this concept by investigating various strategies for fine-tuning Verilog-writing models. More recently, two studies delved into LLMs' applications in chip design. The former, Chip-Chat \cite{refC7}, employed the latest LLM to design an 8-bit shift register chip; while the latter, ChipGPT \cite{refC8}, focused on the power-performance-area (PPA) optimization of an LLM-composed chip design.

In contrast to the aforementioned works, this paper makes significant contributions from two key aspects.

In comparison to the first category of works \cite{refC1,refC2,refC3,refC4}, the first contribution of this paper involves using LLMs to assist FPGA development and offers a comprehensive study of the role of LLMs in facilitating the entire FPGA development process, beyond merely bug fixing. We investigate previously unexplored areas, including validation and maintenance issues, and our research includes the first LLM study that further refines a real-time communication system, OpenWiFi, that has already undergone rigorous validation and practical demonstrations previously.

In comparison to the second category of works \cite{refC5,refC6,refC7,refC8}, our second contribution focuses on utilizing LLMs to write HDL code specifically for wireless communication hardware, involving significantly more complex signal-processing algorithms than those addressed in prior research. Earlier LLM-written HDL codes, as presented in \cite{refC5} and \cite{refC6}, were confined to simple signal processing no harder than undergraduate-level assignments, as acknowledged by the authors in their subsequent work \cite{refC7}. Although \cite{refC7} and \cite{refC8} advanced the complexity by tackling an 8-bit shift register and a multi-functional calculator, these tasks still fall short in complexity compared to communication algorithms such as FFT. As a result of tackling more demanding coding tasks, we encountered two challenges that had not been previously reported in prior works: the subtask scheduling challenge and the multi-step thinking challenge. Addressing these challenges necessitates the use of advanced methods, namely in-context learning (ICL) and chain-of-thought (CoT) prompting, which have never been considered in prior works \cite{refC5,refC6,refC7,refC8}.

%% file: Source/Section_III.tex
This section delves into the various tasks frequently encountered when prototyping wireless systems on FPGAs. Our contribution lies in the proposition of utilizing LLMs to amplify implementation efficiency and productivity in the realm of wireless communication research. By refining OpenWiFi \cite{refB1}, a well-known open-source FPGA-based SDR project, we not only provide valuable insights and practical experiences but also pave the way for AI-assisted SDR development on FPGA. The advantages of harnessing LLMs in hardware development are prominently demonstrated across three pivotal dimensions: 1) Code Refactoring, 2) Code Reuse, and 3) Code Validation.

\subsection{Code Refactoring}
Improving code quality is crucial in FPGA design, as even functioning code may still benefit from further enhancements \cite{ref008}. Code refactoring is a routine task for FPGA engineers, involving manual review and editing. Recent research works suggest that artificial intelligence (AI) can assist in scanning and revising code \cite{ref009}. This subsection demonstrates the competence of LLMs in code refactoring and shows how LLMs can offer valuable assistance to engineers in this kind of work. To illustrate this, we choose a simple signal delay module from OpenWiFi and showcase how an LLM can improve the code in terms of readability, efficiency, and reliability. The original code is presented in Fig. \ref{Fig_1} of Appendix \ref{sec-App1}, while Fig. \ref{Fig_2} of the same appendix presents our prompt that asks ChatGPT to improve the code.

The first objective of code refactoring is to improve its \textit{readability}. A Verilog project that is easy to read and understand facilitates easier future maintenance. Key characteristics of well-written code include consistent programming style, meaningful module/variable names, and sufficient comments that clearly explain code functionality. However, in practice, code contributors collaborating on a project may have their own programming styles. Despite widely accepted programming standards and additional coding requirements within a development team, poorly written code can still present challenges in terms of readability and maintainability.

LLMs offer a consistent programming style, thus aiding in unifying the code within the same FPGA project written by diverse coders. Additionally, LLMs possess the intelligence to address naming and commenting issues effectively. In our experiments, ChatGPT provides readability suggestions from three distinct perspectives. First, ChatGPT recommends using meaningful names for modules and variables. For instance, in Line 1, the original module name ``DelayT" can be replaced with ``DelayBuffer" for clarity. Similarly, in Line 15, variable ``i" can be renamed to ``index" for improved understanding. Second, ChatGPT identifies redundant code within the ``always" block and suggests shortening it to enhance readability. The revised code (presented in Fig. \ref{Fig_3} of Appendix \ref{sec-App1}) demonstrates the enhance of readability. Lastly, ChatGPT automatically adds comments to assist readers in better comprehending the code. We refer the reader to Fig. \ref{Fig_3} of appendix A for more information about these comments.

The second objective of code refactoring is to enhance \textit{efficiency}. Efficiency in hardware language differs significantly from that in software languages like C or Python. In hardware projects, designers must consider the physical implementation of the hardware within the chip after synthesis. For example, as highlighted by ChatGPT, the code within the ``always" block is realized as a shift register in the FPGA. However, this implementation would be wasteful of resources if the function of the IP core is configured to delay the signal for a relatively long time, say more than 10 clock periods. A more efficient approach to implement the delay function is using a read/write counter and a block RAM (mature RAM IP cores are readily available). Note that we do not revise the code based on this suggestion (see Fig. \ref{Fig_3} of appendix A for code details), as it would involve a fundamental redesign of the module. However, we consider this comment to be of significant importance for achieving highly efficient hardware processing in OpenWiFi, especially considering that the delay module is frequently reused in their designs.

The third objective of code refactoring is to enhance \textit{reliability}. Some code may appear to work well in a design simply because the bugs within it are not triggered. For instance, a flawed design may cause problems when the operating voltage or clock frequency is high. Additionally, advanced HDL design tools may automatically correct bugs during the synthesis stage or the place-and-route stage. However, these design tools could not revise the code itself, leaving underlying problems unresolved (although they may not manifest in the final output). Identifying and addressing such problems can be challenging as they produce the correct output at the moment but may cause trouble if triggered in the future, particularly when reusing the code on a different hardware platform with a higher operating voltage or if the HDL design tool changes. Although no one can guarantee bug-free code, and such issues are common in practice (sometimes referred to as ``features" rather than ``bugs"), avoiding such mistakes during the coding stage is crucial for reliable hardware design.

In light of the Verilog code and the prompt provided, ChatGPT highlights two severe problems that could lead to potential system instability. First, ChatGPT suggests adding ``wire" data type specifications to the ``input" and ``output" ports. We consider this comment valuable. Although a Verilog synthesizer can infer the data type of these ports and apply default settings, it is good practice to explicitly state them to be sure. Second, ChatGPT recommends including the negative edge of active-low reset signals (i.e., reset\_n) in the sensitivity list of the ``always" block. This modification offers two advantages: 1) a digital circuitry utilizing an active-low reset signal is less likely to be erroneously triggered by noise compared to those employing an active-high reset signal \cite{ref010}; 2) asynchronous reset is more reliable because the system can respond immediately upon detecting an error, without waiting for the rising edge of the clock signal \cite{ref011}. Based on the author's experience in the IC industry, ``asynchronous active-low reset" is a widely adopted programming standard. And we believe this is a crucial issue overlooked by the developers of OpenWiFi.

In conclusion, through this and similar exercises, we have gained substantial confidence in asserting that LLMs can serve as valuable assistants in improving and refactoring Verilog codes.

\subsection{Code Reuse}
Reusing mature designs is a common approach for efficient FPGA development. Highly configurable code allows for easy reuse by simple parameter adjustments. For example, if we apply the default settings for the Verilog module presented in Fig. \ref{Fig_3} of appendix A, it can delay a 32-bit signal by one clock period. However, if we intend to use the same code for a 64-bit signal and require a delay of four clock periods, we simply need to set the parameters DATA\_WIDTH and DELAY to 64 and 4, respectively.

In realistic engineering scenarios, it is true that not all codes are written in a parameterized manner. This can lead to a large workload when attempting to customize the code for specific requirements. For instance, Fig. \ref{Fig_4} in the appendix shows the complex multiplier module used in OpenWiFi, where the code is specifically designed for input signals with a 16-bit data width. If there is a need to enhance the calculation precision to 32 bits for better signal processing accuracy, it is necessary to invest time to rewrite the code. This process can be tedious and time-consuming, requiring careful modifications and adjustments.

We found that LLMs can provide significant assistance in such tasks. In Fig. \ref{Fig_5} of appendix A, we present the prompt used for the parameterization job, and in Fig. \ref{Fig_6} of the same appendix, we show the new code generated by ChatGPT. As seen, ChatGPT successfully parameterizes the code while maintaining its correctness. The revised code is versatile and capable of accommodating diverse data widths and latency requirements by adjusting the module's parameter settings.

By applying this method to modify the codes in OpenWiFi, we can enhance its user-friendliness for reuse and extensions. This approach facilitates easier customization and adaptation of the project to different specifications and requirements.

\subsection{Code Validation}
In FPGA development, code validation is a routine task that engineers undertake to ensure the correctness of the code. This typically involves writing a testbench and attempting to cover a wide range of input possibilities. This subsection points out a potential shift in the future: we may not need to invest extensive time in testbench development, as LLMs can assist in generating rigorous testbenches with comprehensive input coverage.

In Fig. \ref{Fig_7} of appendix A, we present the prompt used to generate the testbench for the complex multiplier mentioned above. The testbench code generated by ChatGPT is shown in Fig. \ref{Fig_8} of appendix A. It is evident from the code that ChatGPT produces a well-structured testbench, incorporating all the necessary elements. We test the revised complex multiplier with the generated testbench, and the result indicates that both the revised module and the testbench are error-free.

Additionally, ChatGPT can also provide the expected outputs for each input it generates, which can further assist in the code validation process. For example, with the prompt presented in Fig. \ref{Fig_9} of appendix A, we obtain more potential inputs for the testbench, and ChatGPT also outputs the corresponding calculation result for each input. This greatly simplifies the validation process.

%% file: Source/Section_IV.tex
Although previous research \cite{refC5,refC6,refC7,refC8} has demonstrated LLMs' ability in generating basic hardware modules, such as shift registers or dice rollers (as discussed in Section \ref{sec-II}), employing LLMs to generate HDL code for advanced signal-processing algorithms remains unexplored. To bridge this gap, the following two sections present our efforts in pushing the knowledge boundary of LLMs. Specifically, we delve into utilizing LLMs to generate complex HDL code for advanced wireless communication algorithms, going beyond the simple code refinement task discussed in Section \ref{sec-III}. As a case study, we focus on FFT, a complex yet vital component in wireless communication hardware.

In this section, we highlight two challenges when employing LLMs to generate Verilog code for the FFT module, namely the ``subtask scheduling" problem and the ``multi-step thinking" problem. In the next section, we present our approaches to address these challenges through the utilization of in-context learning (ICL) and Chain-of-Thought (CoT) prompting techniques. The analysis in these two sections provides valuable insights to effectively leverage LLMs for generating complex HDL code specifically tailored for \textit{iterative} signal-processing algorithms like FFT.

\subsection{Verilog Code Generated by ChatGPT}
Let us begin with a conversation with ChatGPT. In our experiments, we gave ChatGPT multiple tries and found that ChatGPT was unable to generate the code for a 64-point FFT module (our prompt is given in Fig. \ref{Fig_10} of Appendix A). Specifically, the AI either provided a general framework for the FFT module, requiring additional manual input to complete the specific code, or offered an implementation limited to a trivial two-point FFT. We invite readers to personally engage with this exercise, as it offers a firsthand understanding of the current capabilities of ChatGPT in this particular application.

Upon the failure, we scaled down the complexity of the task to generate an eight-point FFT module instead of a 64-point FFT module. ChatGPT managed to create code that seemed correct and professional (as depicted in Fig. \ref{Fig_11} of Appendix A). However, despite its polished appearance, the generated code failed to function as expected and did not pass the FFT testbench.


\subsection{Challenges One: The Subtask Scheduling Problem}
The first issue we identify in the generated code is that ChatGPT lacks an essential understanding of task scheduling and sequential control.

In general, complex computational tasks can be broken down into a series of simpler subtasks, among which parallel and precedence relationships often exist. For example, in the code provided in Fig. \ref{Fig_11} of Appendix A, we observe that the FFT computation can be decomposed into numerous butterfly computations and complex multiplications. Some subtasks, such as butterfly computations from the same FFT stage, can be executed concurrently. However, other tasks, like butterfly computations of successive stages, have a precedence relationship and cannot be parallelized.

To better elucidate this precedence relationship, we present the flow graph of an eight-point FFT in Fig. \ref{Fig_12} below. Here, it is clear that the output of the orange/blue butterfly computation (in the first FFT stage) serves as the input for the red butterfly computation (in the second FFT stage). We cannot execute the red task until the blue and orange tasks have been completed, indicating a precedence relationship between the red task and the blue/orange task. For a more detailed explanation and formal definition of the precedence relationships among butterfly computations, we refer readers to Section II of a related paper \cite{ref004}.
\begin{figure}[htbp]
  \centering
  \includegraphics[width=0.25\textwidth]{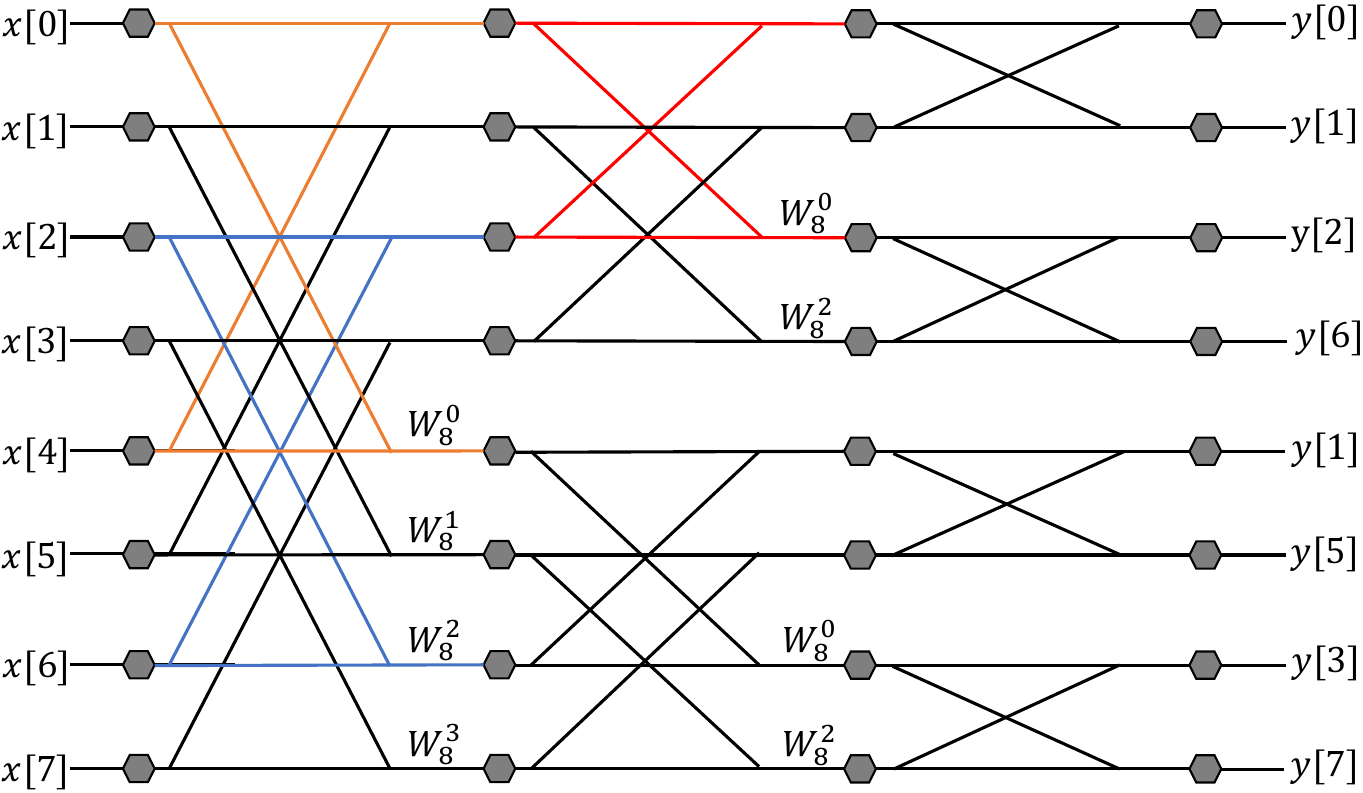}\\
  \captionsetup{font={footnotesize}}
  \caption{The flow graph of an eight-point FFT.}\label{Fig_12}
\end{figure}

Bearing the above discussion in mind, we note that the precedence relationships among butterfly computations necessitate careful subtask scheduling when implementing FFT in Verilog. Unlike high-level programming languages, which have the assistance of compilers and OSs to manage subtask's scheduling\footnote{The scheduling issue can be less complex for high-level programming languages like C or python, as the software compilers and the OS can help to handle the scheduling problem. These tools can distinguish the parallel/precedence relationship and assign the butterfly computations to hardware processers accordingly, and users can just describe their algorithm without too much worry about the scheduling issue. For more details about how to implement FFT in C and what a compiler/OS can help in the implementation, we refer the reader to the documentation of FFTW, a high-performance FFT library \cite{ref012}.}, Verilog faces hardware directly. Therefore, designers need to consider the scheduling issue themselves in the HDL code so that these subtasks can be executed in a sequential manner.

Over the past few decades, FFT implementation has been extensively studied and several classic scheduling schemes have emerged. The simplest approach for precise subtask execution control involves the use of enable signals and output-state-indicating signals. Specifically, if the execution of subtask A depends on the completion of subtask B, we can connect the output-state-indicating signal of B with the enable signal of A to manage their execution. Upon the completion of B, its output indicating signal becomes valid, which subsequently triggers the execution of A. The output-state-indicating signal is sometimes referred to as the ``done" signal, as it becomes valid only when the associated subtask is fully executed.

We now look back to the code generated by ChatGPT. It is apparent that there is no task execution control in the implementation. Neither the basic method of using enable/done signals nor more advanced methods like state machines \cite{ref016} are observed in the code. To confirm our observation, we validated the code using our testbench and found that many subsequent subtasks were prematurely executed before the outputs of their preceding tasks became valid. This resulted in erroneous outputs at the final stage. The experiment results corroborate our initial assertion: ChatGPT, in its current state, lacks awareness of subtask scheduling and sequential execution control. Therefore, it is incapable of generating a viable FFT module autonomously.

In Section \ref{sec-V}-A, we detail our approach to enabling ChatGPT to comprehend the concept of precedence relationships among subtasks and subsequently implement execution control using enable/done signals.

\subsection{Challenges Two: The Multi-step Thinking Problem}
The second issue we identified with the LLM-generated code is the inability of ChatGPT to correctly generate the twiddle factors, a crucial component in FFT calculations (refer to lines 15 to 18 in Fig. \ref{Fig_11} of Appendix A). This issue persisted regardless of the number of attempts or variations in the prompts we used.

Before diving into why ChatGPT is unable to generate these factors, it is necessary to provide a detailed understanding for the concept of twiddle factors. As observed in Fig. \ref{Fig_12}), the data in the course of the algorithm is multiplied by trigonometric constant coefficients, denoted as $W_{N}^{k}={{e}^{-j\left( {2\pi k}/{N} \right)}}$, where $N=8$ is the size of FFT, and index $k\in \left\{ 0,1,...{N}/{2} \right\}$. These coefficients are referred to as the twiddle factors.

In theory, the real and imaginary parts of $W_{N}^{k}$ are numbers no larger than one. In hardware processing, however, things are different because digital circuitry is designed to handle integers expressed in binary form. Here, we illustrate how a human engineer would transform the complex number $W_{N}^{k}$ into a 32-bit binary sequence, with a 16-bit imaginary part and a 16-bit real part, using $W_{8}^{1}$ as an example:
\begin{enumerate}
    \item Step One (calculation): we have $W_{8}^{1}={{e}^{-j\left( {\pi }/{4}\; \right)}}=0.7071-0.7071i$ form trigonometric calculations.
    \item Step Two (scaling): we scale the real and imaginary parts of $W_{8}^{1}$ by multiplying them by a scaling factor, typically chosen as the maximum value that can be represented by the number of bits allocated for each part (in this case, 16 bits). Hence, we amplify $\operatorname{Re}\left( W_{8}^{1} \right)$ and $\operatorname{Im}\left( W_{8}^{1} \right)$ by ${{2}^{15}}-1$. Now, we have $\operatorname{Re}\left( W_{8}^{1} \right)={23169}{.5457}$ and $\operatorname{Im}\left( W_{8}^{1} \right)=-{23169}{.5457}$.
    \item Step Three (rounding): we do rounding operation on $\operatorname{Re}\left( W_{8}^{1} \right)$ and $\operatorname{Im}\left( W_{8}^{1} \right)$, and now we have $\operatorname{Re}\left( W_{8}^{1} \right)\approx {23170}$ and $\operatorname{Im}\left( W_{8}^{1} \right)\approx -23170$.
    \item Step Four (Conversion to binary): we convert $\operatorname{Re}\left( W_{8}^{1} \right)$ and $\operatorname{Im}\left( W_{8}^{1} \right)$ to their binary representations, which is ``0101,1010,1000,0010" and ``1010,0101,0111,1110", respectively.
    \item Step Five (Concatenation): we concatenate the binary representations of the scaled real and imaginary parts to form a 32-bit binary sequence, with the higher 16 bits being the imaginary part and the lower 16 bits being the real part. We can now represent $W_{8}^{1}$ by ``1010,0101,0111,1110, 0101,1010,1000,0010".
\end{enumerate}

By following the above five steps, a human engineer can transform $W_{N}^{k}$ into a 32-bit binary sequence suitable for hardware processing in digital circuitry. One more thing we note is that, when employing the 32-bit sequence for complex multiplication, we need to shrink the multiplication output appropriately to maintain accuracy, as we have amplified $W_{N}^{k}$ in Step Two.

From the previous discussion, it becomes evident that generating the twiddle factors is not a straightforward process. It involves five different steps. Although the logical reasoning required for each individual step might not pose a significant challenge for ChatGPT, the entire problem becomes very difficult for the AI model, as it lacks the ability to decompose the problem into intermediate steps as a human engineer would do. This limitation, known as the lack of multi-step thinking ability, has also been observed in recent research within the NLP community \cite{ref006,ref007}. A number of studies have been carried out to enhance the capabilities of large language models like ChatGPT by aiding them in emulating human-like multi-step reasoning processes\cite{ref005,ref017}. This line of research aims to help AI overcome complex problems that require intermediate steps for solution.

Given the analysis above, we have identified the underlying reason why ChatGPT could not generate the twiddle factors in our initial trials. In Section \ref{sec-V}-B, we will further discuss our approach to addressing this ``multi-step thinking" problem and making ChatGPT able to perform our task.

%% file: Source/Section_V.tex
\subsection{In-context Learning (ICL) for Challenge One}
\textbf{A brief introduction about ICL}

Let us first briefly introduce the concept of ICL. The concept of ICL was popularized in \cite{refA4}, which introduced how to enable GPT-3 to learn from a few examples. In ICL, we give an LLM a prompt containing several question-answer pairs as examples to demonstrate how to complete a task. Following these pairs, a new, unaddressed question is appended to the prompt. The aim is for the LLM to analyze the previously given examples, extrapolate the underlying task, and provide an answer to this new question based on that learning context.

In Fig. \ref{Fig_13} below, we give an example prompt for using LLMs in a news classification task. As input examples, we provide several news titles and their corresponding topic classifications, creating a series of question-answer pairs. We then present the LLM with a news title for which it must generate the relevant topic classification. To correctly answer this question, the model must analyze the provided examples to understand several aspects of the problem: the structure of the input (news titles), the range of possible outputs (possible news topics), the mapping from input to output (topic classification), and the formatting of the output (a single word with the first letter capitalized). With this understanding, ChatGPT generates the correct answer, i.e., ``Technology".

\begin{figure}[htbp]
\centering
\begin{tcolorbox}[sharp corners, colback=white, colframe=black, left=2.5mm]
\begin{lstlisting}[style={prompt-style}]
I want you to help me with a topic classification task. I will give you
examples and you should learn from them. In each example, I give you the
title of a piece of news, and I will tell you the correct topic for the news.

Here are four examples.
  Title  1: CPI in HK averaged 65.72 points from 1980 until 2023.
  Answer 1: Finance

  Title  2: Jonas won le tour de france in 2022.
  Answer 2: Sport
  
  Title  3: Apple developed a new technology called Apple Vision Pro.
  Answer 3: Technology 

  Title  4: Biden announces bid for his second term as US President
  Answer 4: Politics 

Now, here is a piece of news, tell me its classification
Question: Qualcomm announces a new collaboration with Sony to deliver 
          next-generation smartphones.
\end{lstlisting}
\end{tcolorbox}
\captionsetup{font={footnotesize}}
\caption{An ``new classification" example for ICL.}\label{Fig_13}
\end{figure}

ICL distinguishes itself from conventional machine learning algorithms in several key ways \cite{ref018}. Most notably, it does not require any parameter optimization or the addition of new parameters to the model. ICL works effectively with only a handful of training examples to get an LLM operational on a new topic, and its natural language interface is intuitive, even for beginners.

There have been recent research efforts aiming to decipher why ICL performs so remarkably well. The prevailing theory is that an LLM can more effectively ``locate" a previously learned concept with the assistance of ICL. Specifically, since an LLM is trained on a vast amount of text encompassing a wide range of topics and formats, it can model a diverse array of learned concepts with knowledge from various domains. An LLM can deliver better results if we assist it in selecting the most suitable domain knowledge with the hints provided in our ICL examples. For instance, in this paper, our task necessitates greater domain knowledge in HDL, as opposed to languages like C or Python. For a more comprehensive understanding of the underlying mechanisms that make ICL effective, we refer the reader to \cite{ref021}.

\textbf{ICL for our Verilog-writing task}

\begin{figure}[htbp]
  \centering
  \includegraphics[width=0.25\textwidth]{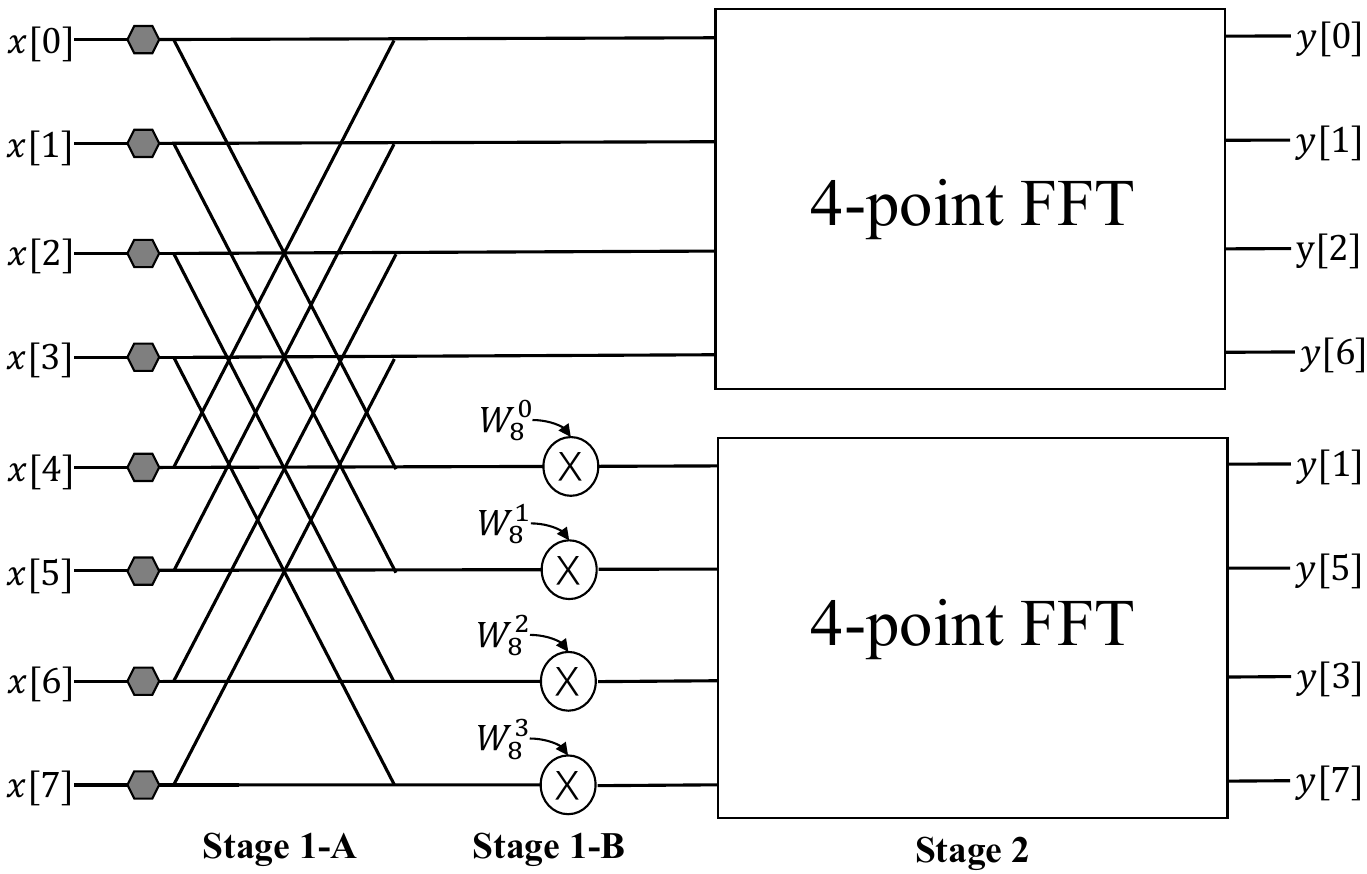}\\
  \captionsetup{font={footnotesize}}
  \caption{The iterative flow graph of an eight-point FFT.}\label{Fig_14}
\end{figure}

We now demonstrate how we use ICL to build the 64-point FFT module in Verilog. We start by re-shaping the FFT flow graph in an iterative manner for ChatGPT's easier understanding and imitation. In a typical FFT flow graph, such as the one presented in Fig. \ref{Fig_12}, an $N$-point FFT has ${{\log }_{2}}N$ stages. In essence, the signal processing in the subsequent ${{\log }_{2}}N-1$ stages can be perceived as two parallel ${N}/{2}$-point FFT processes. Therefore, as illustrated in Fig. \ref{Fig_14}, using an eight-point FFT example, we can simplify the flow graph into two stages: the first stage consists of ${N}/{2}$ butterfly computations and ${N}/{2}$ complex multiplications, we term these two substages as stage 1-A and stage 1-B, respectively. The second stage encompasses two parallel ${N}/{2}$-point FFT processes. With the new iterative flow graph, we simplify the understanding of FFT and aid in the better comprehension of LLMs.

We then analyze the precedence/parallel relationship within the iterative FFT flow graph. It is important to note that, beyond the structure of the flow graph, the available hardware resources can also influence these relationships. For instance, if an FPGA has limited hardware resources, to the point that it can only execute one butterfly computation at a time, the butterfly computations within the same stage (such as the four butterfly computations in Stage 1-A of Fig. \ref{Fig_14}) would have a precedence relationship. This is because they must be executed sequentially, i.e., one after another. On the other hand, if the FPGA holds abundant hardware resources, the butterfly computations within the same stage can be executed in a fully parallel manner. In this paper, we consider a scenario where the FPGA has ample hardware resources so that the precedence/parallelism relationships are solely determined by the flow graph itself. With this assumption, we characterize the precedence/parallelism relationship of an $N$-point FFT as follows:
\begin{enumerate}
    \item \textbf{Stage 1-A}: The ${N}/{2}$ butterfly computations with this stage can be executed in parallel. These computations can be processed simultaneously when triggered by the external enable signal.
    \item \textbf{Stage 1-B}: The ${N}/{2}$ complex multiplications within this stage can be executed in parallel, but their executions are triggered by the completeness of butterfly computations in Stage 1-A.
    \item \textbf{Stage 2} The two ${N}/{2}$-point FFT in Stage 2 within this stage can be executed in parallel, but their executions are triggered by the completeness of complex multiplications in Stage 1-B.
\end{enumerate}

With the simplified precedence/parallel relationships and the iterative FFT flow graph discussed above, we now demonstrate how we generate our question-answer pairs and conduct ICL with the goal of creating a 64-point FFT module using ChatGPT.

In the first step, we use ChatGPT to generate two simple IP cores that will be frequently used in the subsequent FFT implementation: the butterfly computation IP core and the complex multiplication IP core. Fig. \ref{Fig_15} of Appendix A elucidates the prompt specifically devised for this task, while Fig. \ref{Fig_16} and Fig. \ref{Fig_16b} of the same Appendix showcase the code generated by ChatGPT.


In the second step, we give the first question-example pair. The example question, as depicted in Fig. \ref{Fig_17} of Appendix A, asks the LLM to generate a four-point FFT IP core, building upon the provided two-point FFT (which is identical to the butterfly computation IP core). Our example answer, as illustrated in Fig. \ref{Fig_19} of Appendix A, employs two butterfly computations, two complex multiplications, and two two-point FFTs to construct a four-point FFT, adhering to the iterative structure delineated in Fig.  \ref{Fig_14}. Furthermore, this example answer also demonstrates the methodology of connecting ``enable" and ``done" signals of sub-modules to effectuate the precedence/parallel relationships outlined above.

In the third step, we proceed with the second question-example pair. The example question, presented in Fig. \ref{Fig_20} of Appendix A, asks ChatGPT to develop an eight-point FFT module base on the provided four-point FFT, which is obtained in the first question-answer pair. Our example answer, showcased in Fig. \ref{Fig_22} of Appendix A, outlines how the numerous sub-modules (consisting of four butterfly computations, four complex multiplications, and two four-point FFTs) are interconnected in accordance with the iterative FFT flow graph. Furthermore, we present the method of connecting ``enable" and ``done" signals once again, reinforcing this knowledge for ChatGPT.

In the fourth step, we cease providing examples. Instead, we pose a new question to ChatGPT akin to the previous example question: generate a 16-point FFT predicated on the eight-point FFT provided (i.e., the one we give as the example answer in step three). This time, ChatGPT produces an implementation code that is synthesizable and capable of generating outputs identical to those of a Xilinx 16-point FFT IP core, thereby ensuring functional correctness. The only persisting issue pertains to the absence of twiddle factors, a problem that we intend to address in the succeeding subsection (in the above benchmarking with the Xilinx IP core, we filled in the twiddle factors generated in Subsection B to the code).

Finally, we repeat step four in an iterative manner. This is, we ask ChatGPT to generate an $N$-point FFT with the provided ${N}/{2}$-point FFT (which was generated by ChatGPT in the preceding iteration). We do not stop the iteration until we acquire the desired FFT module. In this paper, as a proof of concept, we terminate at the 64-point FFT and present the generated code as in Fig. \ref{Fig_24} of Appendix A. We test the code with our 64-point FFT testbench and compare the output of the LLM-written module with the output of a 64-point FFT IP core provided by Xilinx. Experimental results reveal that the Verilog module, written by ChatGPT, is functionally accurate after the above iterative generation process.

\subsection{Chain-of-Thought (CoT) Prompting for Challenge Two}
\textbf{A brief introduction about CoT prompting}

A typical class of tasks that present challenges to language models is solving mathematical problems, particularly those requiring multi-step reasoning \cite{ref005,ref006,ref007,ref017}. This challenge persisted as a tough problem in the NLP community until the advent of LLMs. In \cite{ref022}, the authors surprisingly discovered that their language model's arithmetic reasoning capability can be dramatically enhanced when the model size scales beyond 100 billion parameters. Furthermore, \cite{ref023} found that guiding an LLM through a human's chain-of-thought in breaking down a multi-step problem into intermediary steps can enable the model in solving complex reasoning problems, which are unattainable with conventional prompting methods. These two inspiring discoveries have inspired a recent surge of research interest in chain-of-thought (CoT) prompting for LLMs.

We now give an example to illustrate the concept of CoT prompting. In this example, as shown in Fig. \ref{Fig_25} below, the baseline prompt comprises ICL with a single example question-answer pair. In contrast, the CoT prompt extends the example answer to incorporate a chain of thought detailing how the problem should be dissected and tackled. For more examples illustrating the efficacy of CoT prompting, we refer interested readers to \cite{ref005}.
\begin{figure}[htbp]
  \centering
  \includegraphics[width=0.45\textwidth]{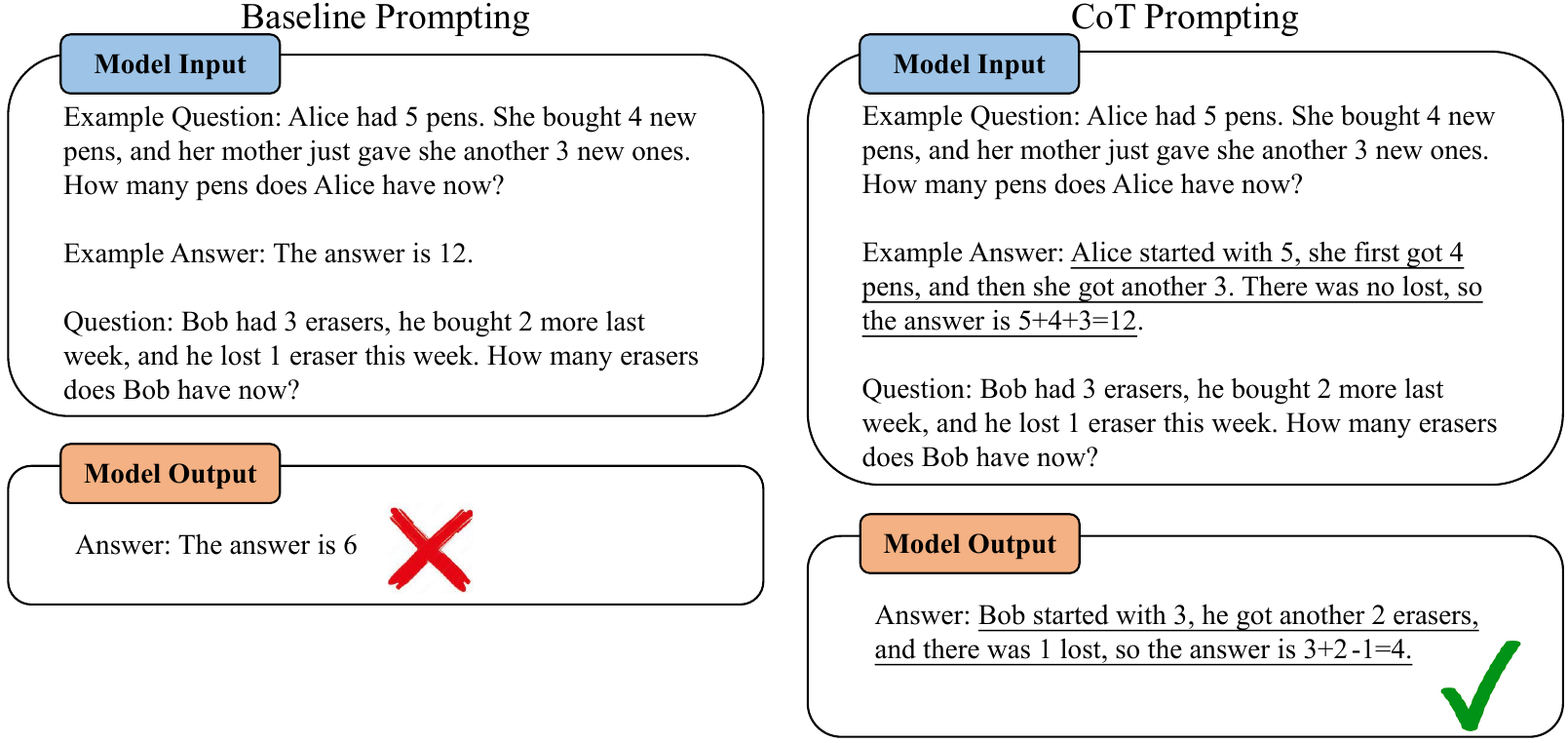}\\
  \captionsetup{font={footnotesize}}
  \caption{An example to illustrate the CoT prompting technique. The CoT reasoning processes are underlined.}\label{Fig_25}
\end{figure}

From the above example, it is evident that ICL coupled with CoT prompting outperforms the baseline approach. However, it is important to note that we do not mean that a contemporary LLM cannot generate the correct answer using the baseline prompt. Our intention is to use this example to demonstrate the concept of CoT prompting and how it should be employed. In fact, LLMs nowadays have advanced beyond those reported in early studies and can produce correct results for the simple question depicted in Fig. \ref{Fig_25}, even without the assistance of ICL or CoT prompting. However, for more complex tasks, such as the twiddle factor generation tasks we describe in Section \ref{sec-IV}-C, we observe that the challenge of multi-step reasoning persists. That motivates us to integrate CoT prompting within the ICL framework in this study.

\textbf{CoT prompting for our twiddle factors generation task}

In Section \ref{sec-IV}-C, we elucidated the process of converting twiddle factors into 32-bit sequences for digital circuitry. Here we describe the multi-step transformation process in detail and design the CoT prompt. As an illustration, our prompt employs the generation process of the twiddle factors for an eight-point FFT (i.e., $W_{8}^{0}$, $W_{8}^{1}$, $W_{8}^{2}$, and $W_{8}^{3}$) as examples. And then we ask ChatGPT to generate the twiddle factors for a 16-point FFT. We depict our prompt and the resulting twiddle factor sequences in Fig. \ref{Fig_26} and Table I, respectively.

\begin{table}[htbp]\label{table:3}
\renewcommand{\arraystretch}{1.25}
\scriptsize
\captionsetup{font={footnotesize}}
\caption{LLM generated 32-bit sequences for twiddle factors in 16-point FFT}
\begin{center}
\begin{tabular}{|p{0.4cm}<{\centering}|p{0.975cm}<{\centering}|p{0.975cm}<{\centering}|p{4.5cm}<{\centering}|}
\hline
&Imag Part & Real Part & 32-bit Sequence \\
\hline
$W_{16}^{0}$& 1.0000 & 0.0000i & 0000,0000,0000,0000, 0111,1111,1111,1111 \\
\hline
$W_{16}^{1}$& 0.9239 & -0.3827i & 1100,1111,0000,0101, 0111,0110,0100,0001 \\
\hline
$W_{16}^{2}$& 0.7071 & -0.7071i & 1010,0101,0111,1110, 0101,1010,1000,0010 \\
\hline
$W_{16}^{3}$& 0.3827 & -0.9239i & 1000,1001,1011,1111, 0011,0000,1111,1011 \\
\hline
$W_{16}^{4}$& 0.0000 & -1.0000i & 1000,0000,0000,0001, 0000,0000,0000,0000 \\
\hline
$W_{16}^{5}$& -0.3827 & -0.9239i & 1000,1001,1011,1111, 1100,1111,0000,0101 \\
\hline
$W_{16}^{6}$& -0.7071 & -0.7071i  & 1010,0101,0111,1110, 1010,0101,0111,1110 \\
\hline
$W_{16}^{7}$& -0.9239 & -0.3827i & 1100,1111,0000,0101, 1000,1001,1011,1111 \\
\hline
\end{tabular}
\end{center}
\end{table}

Moreover, we verify that the same CoT prompt (employing $W_{8}^{0}$, $W_{8}^{1}$, $W_{8}^{2}$, and $W_{8}^{3}$ as examples) is applicable for generating twiddle factors for larger-scale FFTs, such as 32-point or 64-point. In other words, we can skip teaching ChatGPT about the factors generation of a 32-point FFT and directly jump to the factor generation of a 64-point FFT, which affirms that the LLM does internalize the crucial knowledge imparted through the CoT prompt (rather than simply parroting the input). Here we do not present the twiddle factor generation process for larger-scale FFTs due to page limitation. We encourage interested readers to give it a try themselves.

%% file: Source/Section_New_Benchmarking.tex
To quantitatively evaluate the effect of using LLMs in FPGA development, we designed three representative Verilog programming tasks and conducted benchmark experiments among four groups of volunteer students. Experimental results indicate the significant productivity gain brought by LLM on FPGA development. We refer readers to Appendix B of this paper for detailed information about 1) test questions, 2) our evaluation criteria, 3) the distributions and backgrounds of volunteers, 4) experimental results, and 5) discussions and analyses associated with the data.

%% file: Source/Section_VI.tex
This paper delves into the intersection of Large Language Models (LLMs) and wireless communication technologies, yielding inspiring results in utilizing LLMs to prototype wireless systems. Our research highlights the potential of LLMs in facilitating complex FPGA development within wireless systems.

We begin by demonstrating how an LLM can serve as a crucial assistant for FPGA development, providing examples in code refactoring, code reuse, and system validation. Moreover, we showcase LLMs' ability to generate sophisticated Hardware Description Language (HDL) codes for advanced signal-processing algorithms in wireless communication, with a focus on the fundamental Fast Fourier Transform (FFT) processing.

By addressing the subtask scheduling problem and multi-step thinking problem through In-context Learning (ICL) and the Chain of Thoughts (CoT) prompting techniques, we successfully generated a 64-point Verilog FFT module using LLMs for the first time. This exploration of LLMs' generalization and imitation capabilities expands their potential applications and underscores their value in the wireless communication domain.

%% file: Source/Appendix_A1.tex
\setcounter{figure}{0} 
\begin{figure}[htbp]
\centering
\begin{tcolorbox}[sharp corners, colback=white, colframe=black, left=2.5mm]
\begin{lstlisting}[style={verilog-style}]
module delayT #( parameter DATA_WIDTH = 32, parameter DELAY = 1 ) (
  input clock,
  input reset,
  input [DATA_WIDTH-1:0] data_in,
  output [DATA_WIDTH-1:0] data_out
);
  reg [DATA_WIDTH-1:0] ram[DELAY-1:0];
  integer i;
  assign data_out = ram[DELAY-1];
  always @(posedge clock) begin
    if (reset) begin
      for (i = 0; i < DELAY; i = i+1) begin
        ram[i] <= 0;
      end
    end 
    else begin
      ram[0] <= data_in;
      for (i = 1; i < DELAY; i= i+1) begin
        ram[i] <= ram[i-1];
      end
    end
  end
endmodule
\end{lstlisting}
\end{tcolorbox}
\captionsetup{font={footnotesize}}
\caption{The original code of a signal delay module in OpenWiFi.}\label{Fig_1}
\end{figure}

\begin{figure}[htbp]
\centering
\begin{tcolorbox}[sharp corners, colback=white, colframe=black, left=2.5mm]
\begin{lstlisting}[style={prompt-style}]
I am writing a signal delay module for my FPGA project with Verilog language.
Please help me revise the code and make it easier to understand and error-free.
Besides providing the revised code, please also highlight your changes 
and explain why you modify the code in this way.
Here is the original code:......
\end{lstlisting}
\end{tcolorbox}
\captionsetup{font={footnotesize}}
\caption{The prompt we used to improve the signal delay module.}\label{Fig_2}
\end{figure}

\begin{figure}[htbp]
\centering
\begin{tcolorbox}[sharp corners, colback=white, colframe=black, left=2.5mm, breakable]
\begin{lstlisting}[style={verilog-style}]
module delayT
#(
    // These are module parameters, and can be overridden when the module is instantiated.
    parameter DATA_WIDTH = 32,  // The width of the data to be delayed
    parameter DELAY = 10  // The number of clock cycles to delay the data
)
(
    input wire clock,  // The clock signal
    input wire reset,  // Reset signal; when high, clears the delay line
    input wire [DATA_WIDTH-1:0] data_in,  // Input data to be delayed
    output wire [DATA_WIDTH-1:0] data_out  // Output of the delayed data
);

// This is a memory array for storing the data as it's delayed.
// Its size is controlled by the DELAY parameter.
reg [DATA_WIDTH-1:0] ram[0:DELAY-1]; 
integer i;

always @(posedge clock or posedge reset) begin 
    // Reset has priority. When asserted, it will clear the delay line.
    if (reset) begin
        // This loop clears the delay line to zero.
        for (i = 0; i < DELAY; i = i+1) begin
            ram[i] <= 0;
        end
    end 
    // If not reset, shift the data through the delay line.
    else begin
        // This loop shifts the data towards the end of the delay line.
        for (i = DELAY-1; i > 0; i = i - 1) begin 
            ram[i] <= ram[i-1];
        end
        // New data is placed at the beginning of the delay line.
        ram[0] <= data_in; 
    end
end

// The output of the module is the data at the end of the delay line.
assign data_out = ram[DELAY-1]; 

endmodule
\end{lstlisting}
\end{tcolorbox}
\captionsetup{font={footnotesize}}
\caption{The revised code generated by ChatGPT.}\label{Fig_3}
\end{figure}


%% file: Source/Appendix_A2.tex
\begin{figure}[htbp]
\centering
\begin{tcolorbox}[sharp corners, colback=white, colframe=black, left=2.5mm]
\begin{lstlisting}[style={verilog-style}]
module complex_mult (
  input clock,
  input enable,
  input reset,
  input [15:0] a_i,
  input [15:0] a_q,
  input [15:0] b_i,
  input [15:0] b_q,
  input input_strobe,
  output reg [31:0] p_i,
  output reg [31:0] p_q,
  output output_strobe
);
  reg [15:0] ar, ai, br, bi;
  wire [31:0] prod_i, prod_q;
  wire [31:0] s_axis_a_tdata = {ai,ar} ;
  wire [31:0] s_axis_b_tdata = {bi, br} ;
  wire [63:0] m_axis_dout_tdata;

  assign prod_q = m_axis_dout_tdata[63:32];
  assign prod_i = m_axis_dout_tdata[31:0];
  complex_multiplier mult_inst (
    .aclk(clock),
    .s_axis_a_tvalid(input_strobe),
    .s_axis_a_tdata(s_axis_a_tdata),
    .s_axis_b_tvalid(input_strobe),
    .s_axis_b_tdata(s_axis_b_tdata),
    .m_axis_dout_tdata(m_axis_dout_tdata)
  );
  delayT #(.DATA_WIDTH(1), .DELAY(5)) stb_delay_inst (
    .clock(clock),
    .reset(reset),
    .data_in(input_strobe),
    .data_out(output_strobe)
  );
  always @(posedge clock) begin
    if (reset) begin
      ar <= 0; ai <= 0; 
      br <= 0; bi <= 0; 
      p_i <= 0; p_q <= 0;
    end 
    else if (enable) begin
      ar <= a_i; ai <= a_q; 
      br <= b_i; bi <= b_q; 
      p_i <= prod_i; p_q <= prod_q;
    end
  end
endmodule
\end{lstlisting}
\end{tcolorbox}
\captionsetup{font={footnotesize}}
\caption{The original Verilog code for the complex multiplier in OpenWiFi.}\label{Fig_4}
\end{figure}

\begin{figure}[htbp]
\centering
\begin{tcolorbox}[sharp corners, colback=white, colframe=black, left=2.5mm]
\begin{lstlisting}[style={prompt-style}]
I am writing a Verilog module for my FPGA system. I have a complex multiplier 
that deals with 16-bit input signal. However, I want to rewrite the code in a
parameterized manner. You need to use the parameter function in Verilog to make
the data width of a_i,a_q, b_i, b_q, p_i, p_q become configurable. 
Here is the code: ...
\end{lstlisting}
\end{tcolorbox}
\captionsetup{font={footnotesize}}
\caption{The prompt we used to parameterize the complex multiplier.}\label{Fig_5}
\end{figure}

\begin{figure}[htbp]
\centering
\begin{tcolorbox}[sharp corners, colback=white, colframe=black, left=2.5mm]
\begin{lstlisting}[style={verilog-style}]
module complex_mult #(parameter DATA_WIDTH = 16, parameter PRODUCT_WIDTH = 32) (
  input clock,
  input enable,
  input reset,
  input [DATA_WIDTH-1:0] a_i,
  input [DATA_WIDTH-1:0] a_q,
  input [DATA_WIDTH-1:0] b_i,
  input [DATA_WIDTH-1:0] b_q,
  input input_strobe,
  output reg [PRODUCT_WIDTH-1:0] p_i,
  output reg [PRODUCT_WIDTH-1:0] p_q,
  output output_strobe
);
  reg [DATA_WIDTH-1:0] ar, ai, br, bi;
  wire [PRODUCT_WIDTH-1:0] prod_i, prod_q;
  wire [2*DATA_WIDTH-1:0] s_axis_a_tdata = {ai,ar} ;
  wire [2*DATA_WIDTH-1:0] s_axis_b_tdata = {bi, br} ;
  wire [2*PRODUCT_WIDTH-1:0] m_axis_dout_tdata;
  assign prod_q = m_axis_dout_tdata[2*PRODUCT_WIDTH-1:PRODUCT_WIDTH];
  assign prod_i = m_axis_dout_tdata[PRODUCT_WIDTH-1:0];
  complex_multiplier #(2*DATA_WIDTH, 2*PRODUCT_WIDTH) mult_inst (
    .aclk(clock),
    .s_axis_a_tvalid(input_strobe),
    .s_axis_a_tdata(s_axis_a_tdata),
    .s_axis_b_tvalid(input_strobe),
    .s_axis_b_tdata(s_axis_b_tdata),
    .m_axis_dout_tdata(m_axis_dout_tdata)
  );
  delayT #(.DATA_WIDTH(1), .DELAY(5)) stb_delay_inst (
    .clock(clock),
    .reset(reset),
    .data_in(input_strobe),
    .data_out(output_strobe)
  );
  always @(posedge clock) begin
    if (reset) begin
     ar <= 0; ai <= 0; br <= 0; bi <= 0; p_i <= 0; p_q <= 0;
    end 
    else if (enable) begin
      ar <= a_i; ai <= a_q; br <= b_i; bi <= b_q; p_i <= prod_i; p_q <= prod_q;
    end
  end
endmodule
\end{lstlisting}
\end{tcolorbox}
\captionsetup{font={footnotesize}}
\caption{The revised Verilog code for the complex multiplier.}\label{Fig_6}
\end{figure}

%% file: Source/Appendix_A3.tex
\begin{figure}[htbp]
\centering
\begin{tcolorbox}[sharp corners, colback=white, colframe=black, left=2.5mm]
\begin{lstlisting}[style={prompt-style}]
Please write a testbench for the code you wrote. I want you to use a 32-bit data
width. And I want you to test at least ten possible inputs in your testbench.
\end{lstlisting}
\end{tcolorbox}
\captionsetup{font={footnotesize}}
\caption{The prompt we used to generate the testbench for the revised complex multiplier module.}\label{Fig_7}
\end{figure}

\begin{figure}[htbp]
\centering
\begin{tcolorbox}[sharp corners, colback=white, colframe=black, left=2.5mm]
\begin{lstlisting}[style={verilog-style}]
`timescale 1ns / 1ps
module tb_complex_mult;
  reg clock;
  reg enable;
  reg reset;
  reg [31:0] a_i, a_q, b_i, b_q;
  reg input_strobe;
  wire [63:0] p_i, p_q;
  wire output_strobe;
// Instantiate the complex_mult module
  complex_mult #(32, 64) u1 (
    .clock(clock),
    .enable(enable),
    .reset(reset),
    .a_i(a_i),
    .a_q(a_q),
    .b_i(b_i),
    .b_q(b_q),
    .input_strobe(input_strobe),
    .p_i(p_i),
    .p_q(p_q),
    .output_strobe(output_strobe)
  );
// Generate a clock signal
  always #5 clock = ~clock;
// Test sequence
  initial begin
    // Reset
    clock = 0;
    reset = 1; enable = 0; input_strobe = 0;
    // Initialize
    #10; a_i = 0; a_q = 0; b_i = 0; b_q = 0;
    reset = 0; enable = 1; input_strobe = 1;
    // Apply test inputs
    #10; a_i = 10; a_q = 20; b_i = 30; b_q = 40; 
    #10; a_i = 50; a_q = 60; b_i = 70; b_q = 80;
    #10; a_i = 90; a_q = 100; b_i = 110; b_q = 120; 
    #10; a_i = 130; a_q = 140; b_i = 150; b_q = 160; 
    #10; a_i = 170; a_q = 180; b_i = 190; b_q = 200; 
    #10; a_i = 210; a_q = 220; b_i = 230; b_q = 240; 
    #10; a_i = 250; a_q = 260; b_i = 270; b_q = 280; 
    #10; a_i = 290; a_q = 300; b_i = 310; b_q = 320; 
    #10; a_i = 330; a_q = 340; b_i = 350; b_q = 360; 
    #10; a_i = 370; a_q = 380; b_i = 390; b_q = 400; 
    // Finish the test
    $finish;
  end
endmodule
\end{lstlisting}
\end{tcolorbox}
\captionsetup{font={footnotesize}}
\caption{The testbench generated by ChatGPT.}\label{Fig_8}
\end{figure}

\begin{figure}[htbp]
\centering
\begin{tcolorbox}[sharp corners, colback=white, colframe=black, left=2.5mm]
\begin{lstlisting}[style={prompt-style}]
Please give me more possible input for the testbench. 
You don't need to re-write the testbench, just give me the input. 
Further, I want you to tell me the expected output for each possible input.
\end{lstlisting}
\end{tcolorbox}
\captionsetup{font={footnotesize}}
\caption{The prompt we used to generate more inputs for the testbench.}\label{Fig_9}
\end{figure}

%% file: Source/Appendix_A4.tex
\begin{figure}[htbp]
\centering
\begin{tcolorbox}[sharp corners, colback=white, colframe=black, left=2.5mm]
\begin{lstlisting}[style={prompt-style}]
Help me write an FFT module for my FPGA system in Verilog language.
Here are details of my specifications:

  1. The FFT unit should have a length of 64 points. 
  2. The input signal has 16 bits in the real part and 
     16 bits in the imaginary part
  3. The output of the FFT should be of natural order
  4. The target input clock frequency is 200MHz

I also provide you with the instantiation templet:
  fft_GPT_Generate your_instance_name (
    .clk(aclk),
    .resetn(resetn),
    .enable(enable),
    .FFI_input(FFI_input),
    .input_valid(input_valid),
    .FFT_output(FFT_output),
    .output_valid(output_valid)
  );
\end{lstlisting}
\end{tcolorbox}
\captionsetup{font={footnotesize}}
\caption{The prompt we used to generate a 64-point FFT.}\label{Fig_10}
\end{figure}


\begin{figure}[htbp]
\centering
\begin{tcolorbox}[sharp corners, colback=white, colframe=black, left=2.5mm]
\begin{lstlisting}[style={verilog-style}]
module fft_GPT_generate_8_point (
  input wire clk,
  input wire resetn,
  input wire enable,
  input wire [31:0] FFI_input_1,
......// Author: Skip to save space
  input wire [31:0] FFI_input_8,
  input wire input_valid,
  output wire [31:0] FFT_output_1,
......// Author: Skip to save space
  output wire [31:0] FFT_output_8,
  output wire output_valid
);
\end{lstlisting}
\end{tcolorbox}
\end{figure}
\begin{figure}[htbp]
\centering
\begin{tcolorbox}[sharp corners, colback=white, colframe=black, left=2.5mm]
\begin{lstlisting}[firstnumber=14, style={verilog-style}]
// Please fill in the Twiddle Factor
  localparam Twiddle_Factor_8_0=...
  localparam Twiddle_Factor_8_1=...
  localparam Twiddle_Factor_8_2=...
  localparam Twiddle_Factor_8_3=...

// Realize a butterfly computation submodule 
  module butterfly (input wire [31:0] X1, input wire [31:0] X2,
                    output wire [31:0] Y1, output wire [31:0] Y2);
		......// Author: Skip to save space
  endmodule
  
// Realize a complex multiplication submodule 
  module comp_multi (input wire [31:0] X1, input wire [31:0] X2,
                     output wire [31:0] Y);
		......// Author: Skip to save space
  endmodule

//Internal signals
  wire [31:0] o11,o12,o13,o14,o15,o16,o17,o18;
  wire [31:0] o15_shift,o16_shift,o17_shift,o18_shift;
  wire [31:0] o21,o22,o23,o24,o25,o26,o27,o28;  
  wire [31:0] o23_shift,o24_shift,o27_shift,o28_shift;

// Connections of submodules
  butterfly butterfly_1(.X1(FFI_input_1), .X2(FFI_input_5), .Y1(o11), .Y2(o15));
  butterfly butterfly_2(.X1(FFI_input_2), .X2(FFI_input_6), .Y1(o12), .Y2(o16));
  butterfly butterfly_3(.X1(FFI_input_3), .X2(FFI_input_7), .Y1(o13), .Y2(o17));
  butterfly butterfly_4(.X1(FFI_input_4), .X2(FFI_input_8), .Y1(o14), .Y2(o18));
  comp_multi comp_multi_1(.X1(o15), .X2(Twiddle_Factor_8_0), .Y(o15_shift));
  comp_multi comp_multi_2(.X1(o16), .X2(Twiddle_Factor_8_1), .Y(o16_shift));
  comp_multi comp_multi_3(.X1(o17), .X2(Twiddle_Factor_8_2), .Y(o17_shift));
  comp_multi comp_multi_4(.X1(o18), .X2(Twiddle_Factor_8_3), .Y(o18_shift));
  butterfly butterfly_5(.X1(o11), .X2(o13), .Y1(o21), .Y2(o23));
  butterfly butterfly_6(.X1(o12), .X2(o14), .Y1(o22), .Y2(o24));
  butterfly butterfly_7(.X1(o15_shift), .X2(o17_shift), .Y1(o25), .Y2(o27));
  butterfly butterfly_8(.X1(o16_shift), .X2(o18_shift), .Y1(o26), .Y2(o28));
  comp_multi comp_multi_5(.X1(o23), .X2(Twiddle_Factor_8_0), .Y(o23_shift));
  comp_multi comp_multi_6(.X1(o24), .X2(Twiddle_Factor_8_2), .Y(o24_shift));
  comp_multi comp_multi_7(.X1(o27), .X2(Twiddle_Factor_8_0), .Y(o27_shift));
  comp_multi comp_multi_8(.X1(o28), .X2(Twiddle_Factor_8_2), .Y(o28_shift));
  butterfly butterfly_9(.X1(o21), .X2(o22), 
                        .Y1(FFT_output_1), .Y2(FFT_output_2));
  butterfly butterfly_10(.X1(o23_shift), .X2(o24_shift), 
                         .Y1(FFT_output_3), .Y2(FFT_output_4));
  butterfly butterfly_11(.X1(o25), .X2(o26), 
                         .Y1(FFT_output_5), .Y2(FFT_output_6));
  butterfly butterfly_12(.X1(o27_shift), .X2(o28_shift), 
                         .Y1(FFT_output_7), .Y2(FFT_output_8));
endmodule
\end{lstlisting}
\end{tcolorbox}
\captionsetup{font={footnotesize}}
\caption{The eight-point FFT code produced by ChatGPT. To conserve space, we omit portions of the code that are highly similar or repetitive (refer to our comments on Line 6, Line 10, Line 23, and Line 29 for more details).}\label{Fig_11}
\end{figure}

%% file: Source/Appendix_A5.tex
\begin{figure}[htbp]
\centering
\begin{tcolorbox}[sharp corners, colback=white, colframe=black, left=2.5mm]
\begin{lstlisting}[style={prompt-style}]
I am working on an FPGA project in Verilog. Please write two IP cores for me.

The first IP core is for butterfly computation for FFT. Here is its template:
  butterfly_computation your_inst_name( 
    .clk(clk),
    .resetn(resetn),
    .enable(enable),
    .in1(in1),
    .in2(in2),
    .done(done),
    .out1(out1),
    .out2(out2)
  );

The second IP core is for complex multiplication in FFT. I will use it to 
multiply the output of a butterfly computation with the twiddle factor 
provided. Please note that the twiddle factor is amplified by 2^15, 
you need to shrink the out accordingly.

Here is a template of the IP core, 
  butterfly_computation your_inst_name( 
    .clk(clk), 
    .resetn(resetn),
    .enable(enable),
    .in1(in1),
    .in2(in2),
    .done(done),
    .out(out)
  );
\end{lstlisting}
\end{tcolorbox}
\captionsetup{font={footnotesize}}
\caption{The prompt we used to generate the butterfly computation IP core and the complex multiplication IP core.}\label{Fig_15}
\end{figure}

\begin{figure}[htbp]
  \centering
\begin{tcolorbox}[sharp corners, colback=white, colframe=black, left=2.5mm]
\begin{lstlisting}[style={verilog-style}]
module butterfly_computation (
  input wire clk,
  input wire resetn,
  input wire enable,
  input wire [31:0] in1,
  input wire [31:0] in2,
  output reg done,
  output wire [31:0] out1, 
  output wire [31:0] out2
);
// Separate the real and imag parts
  wire [15:0] in1_real = in1[15:0];
  wire [15:0] in1_imag = in1[31:16];
  wire [15:0] in2_real = in2[15:0];
  wire [15:0] in2_imag = in2[31:16];
// Assign the output
  reg [15:0] temp1_real, temp1_imag;
  reg [15:0] temp2_real, temp2_imag;
  assign out1 = {temp1_imag, temp1_real};
  assign out2 = {temp2_imag, temp2_real};
// Alwyas block for assignment
  always @(posedge clk or negedge resetn) begin
    if (~resetn) begin
      done <= 1'b0;
      temp1_real <= 16'd0; temp1_imag <= 16'd0;
      temp2_real <= 16'd0; temp2_imag <= 16'd0;
    end
    else if (enable) begin
      temp_real1 <= in1_real + in2_real;
      temp_imag1 <= in1_imag + in2_imag;
      temp_real2 <= in1_real - in2_real;
      temp_imag2 <= in1_imag - in2_imag;
      done <= 1'b1;
    end
    else
      done <= 1'b0;
  end
endmodule
\end{lstlisting}
\end{tcolorbox}
\captionsetup{font={footnotesize}}
\caption{The Verilog code generated by GPT upon the prompt. Implementation details of the butterfly computation module is presented.}\label{Fig_16}
\end{figure}

\begin{figure}[htbp]
  \centering
\begin{tcolorbox}[sharp corners, colback=white, colframe=black, left=2.5mm]
\begin{lstlisting}[style={verilog-style}]
module complex_multiply (
  input wire clk,
  input wire resetn,
  input wire enable,
  input wire [31:0] in1,
  input wire [31:0] in2,
  output reg done,
  output wire [31:0] out
);
// Separate the real and imag parts
  wire [15:0] in1_real = in1[15:0];
  wire [15:0] in1_imag = in1[31:16];
  wire [15:0] in2_real = in2[15:0];
  wire [15:0] in2_imag = in2[31:16];
// Registers for storing intermediate products
  reg [31:0] real_product;
  reg [31:0] imag_product;
  reg [31:0] x1, x2, x3, x4;
  assign x1 = in1_real * in2_real;
  assign x2 = in1_imag * in2_imag;
  assign x3 = in1_real * in2_imag;
  assign x4 = in1_imag * in2_real;  
  assign out[15:0] = real_product;
  assign out[31:16] = imag_product;
// Always block to assign intermediate products
  always @(posedge clk or negedge resetn) begin
    if (~resetn) begin
      done <= 1'b0;
      real_product <= 32'd0;
      imag_product <= 32'd0;
    end
    else if (enable) begin
      // Shrink real_product and imag_product by 2^15
      real_product <= (x1 - x2) >>> 15; 
      imag_product <= (x3 + x4) >>> 15;
      done <= 1'b1;
    end
    else
      done <= 1'b0;
  end
endmodule
\end{lstlisting}
\end{tcolorbox}
\captionsetup{font={footnotesize}}
\caption{The Verilog code generated by GPT upon the prompt. Implementation details of the complex multiplication module are presented.}\label{Fig_16b}
\end{figure}

\begin{figure}[htbp]
\centering
\begin{tcolorbox}[sharp corners, colback=white, colframe=black, left=2.5mm]
\begin{lstlisting}[style={prompt-style}]
I am writing a four-point DIF-FFT on FPGA. You can use the following IP cores 
to build the target four-point FT IP core.

Here is the template of butterfly computation IP Core,
  butterfly_computation butterfly_inst(
    .clk(),
    .reset(),
    .enable(),
    .in1(), .in2(),
    .done(), 
    .out1(), .out2()
  );
\end{lstlisting}
\end{tcolorbox}
\end{figure}
\begin{figure}[htbp]
\centering
\begin{tcolorbox}[sharp corners, colback=white, colframe=black, left=2.5mm]
\begin{lstlisting}[firstnumber=13, style={prompt-style}]
And here is the template of the two-point FFT IP Core,
  fft_2_point fft_2_point_inst(
    .clk(),
    .reset(),
    .enable(),
    .x1(), .x2(),
    .done(),
    .y1(), .y2()
  );
And here is the template of complex multiplication IP Core,
  complex_multiplication complex_multiplication_inst(
    .clk(),
    .reset(),
    .enable(),
    .in1(), .in2(),
    .done(),
    .out()
  );
  
I also give you the template of the target four-point FFT IP Core,
  fft_4_point fft_4_point_stage_2_1(
    .clk(),
    .reset(),
    .enable(),
    .x1(), .x2(), .x3(), .x4(),
    .y1(), .y2(), .y3(), .y4(),
    .done() 
  );

Further, I also have some suggestions for you.

First, you don't need to calculate the expressions of twiddle factors.
I will fill in the value myself later.

Second, let me tell you the major spirit in developing the 4-point FFT. 
In the first stage of FFT, you have two substages called Stage 1a and 1b. 
In stage 1a, you use twobutterfly_computation modules; and then in stage 1b, 
you use two twiddle models. In the second stage, you are facing a two 
two-point FFT problem, so you just use two fft_2_point and solve the problem.
\end{lstlisting}
\end{tcolorbox}
\captionsetup{font={footnotesize}}
\caption{Question No.1: to generate a four-point FFT IP core with the provided two-point FFT IP core.}\label{Fig_17}
\end{figure}

\begin{figure}[htbp]
\centering
\begin{tcolorbox}[sharp corners, colback=white, colframe=black, left=2.5mm]
\begin{lstlisting}[style={verilog-style}]
module fft_4_point (
  input wire clk,
  input wire reset,
  input wire enable,
  input wire [31:0] x1, input wire [31:0] x2,
  input wire [31:0] x3, input wire [31:0] x4,
  output wire [31:0] y1, output wire [31:0] y2,
  output wire [31:0] y3, output wire [31:0] y4,
  output wire done
);

// Please fill in the expression of twiddle factors
  localparam [31:0] Twiddle_Factor_W_4_0 = ...;
  localparam [31:0] Twiddle_Factor_W_4_1 = ...;

//Stage 1a
  wire stage_1a_done;
  wire stage_1a_enable = enable;
  wire [31:0] stage_1a_out1, stage_1a_out2;
  wire [31:0] stage_1a_out3, stage_1a_out4;
	
  butterfly_computation butterfly_stage_1a_1( 
    .clk(clk), 
    .reset(reset), 
    .enable(stage_1a_enable), 
    .in1(x1), .in2(x3), 
    .out1(stage_1a_out1), .out2(stage_1a_out3), 
    .done() );
				
  butterfly_computation butterfly_stage_1a_2( 
    .clk(clk), 
    .reset(reset), 
    .enable(stage_1a_enable), 
    .in1(x2), .in2(x4), 
    .out1(stage_1a_out2), .out2(stage_1a_out4), 
    .done(stage_1a_done) );
//Stage 1b
  wire stage_1b_done;
  wire stage_1b_enable = stage_1a_done;
  wire [31:0] stage_1b_out1, stage_1b_out2;
  wire [31:0] stage_1b_out3, stage_1b_out4;
	
  assign stage_1b_out1 = stage_1a_out1;
  assign stage_1b_out2 = stage_1a_out2;
	
  complex_multiplication twiddle_stage_1b_1( 
    .clk(clk),
    .reset(reset),
    .enable(stage_1b_enable),
    .in1(stage_1a_out3),.in2(Twiddle_Factor_W_4_0),
    .out(stage_1b_out3),
    .done() );
				
  complex_multiplication twiddle_stage_1b_2( 
    .clk(clk),
    .reset(reset),
    .enable(stage_1b_enable),
    .in1(stage_1a_out4),.in2(Twiddle_Factor_W_4_1),
    .out(stage_1b_out4),
    .done(stage_1b_done) );
\end{lstlisting}
\end{tcolorbox}
\end{figure}

\begin{figure}[htbp]
\centering
\begin{tcolorbox}[sharp corners, colback=white, colframe=black, left=2.5mm]
\begin{lstlisting}[firstnumber=61, style={verilog-style}]   
//Stage 2
  wire stage_2_done;
  assign stage_2_enable = stage_1b_done;
  wire [31:0] stage_2_out1, stage_2_out2;
  wire [31:0] stage_2_out3, stage_2_out4;
	
  fft_2_point fft_2_point_stage_2_1( 
    .clk(clk), 
    .reset(reset), 
    .enable(stage_2_enable), 
    .x1(stage_1b_out1), .x2(stage_1b_out2), 
    .y1(stage_2_out1), .y2(stage_2_out2),
    .done() );
				
  fft_2_point fft_2_point_stage_2_2( 
    .clk(clk), 
    .reset(reset), 
    .enable(stage_2_enable), 
    .x1(stage_1b_out3), .x2(stage_1b_out4), 
    .y1(stage_2_out3), .y2(stage_2_out4),
    .done(stage_2_done) );

//Final output
  assign y1 = stage_2_out1;
  assign y2 = stage_2_out2;
  assign y3 = stage_2_out3;
  assign y4 = stage_2_out4;
  assign done = stage_2_done;
endmodule
\end{lstlisting}
\end{tcolorbox}
\captionsetup{font={footnotesize}}
\caption{Example No.1: a functional four-point FFT generated by ChatGPT.}\label{Fig_19}
\end{figure}

\begin{figure}[htbp]
  \centering
\begin{tcolorbox}[sharp corners, colback=white, colframe=black, left=2.5mm]
\begin{lstlisting}[style={prompt-style}]
I am writing an eight-point DIF-FFT on FPGA. Apart from IP cores Given 
in Question One, i.e., butterfly_computation, fft_2_point, and complex
multiplication, you can also use the fft_4_point IP core generated in
Answer one. You need to look back to Question-1 and Answer-1 for 
detailed input/output information on the four IP cores.

Once again, I want to emphasize that: 
1) no need to calculate the twiddle factors, as  I will fill in the
value later. And 2) let me tell you the major spirit in developing
the 8-point FFT. In the first stage of FFT, you have two substages 
called Stage 1a and 1b. In stage 1a, you use four butterfly_computation
modules; and then in stage 1b, you use four twiddle modules. In the second
stage, you are facing two four-point FFT problems, so you just use two
fft_4_point and solve the problem.
\end{lstlisting}
\end{tcolorbox}
\captionsetup{font={footnotesize}}
\caption{Question No.2: to generate an eight-point FFT IP core with the provided four-point FFT IP core.}\label{Fig_20}
\end{figure}


\begin{figure}[htbp]
\centering
\begin{tcolorbox}[sharp corners, colback=white, colframe=black, left=2.5mm]
\begin{lstlisting}[style={verilog-style}]
module fft_8_point (
  input wire clk,
  input wire reset,
  input wire enable,
  input wire [31:0] x1,
  ......// Author: Skip to save space
  input wire [31:0] x8,
  output wire [31:0] y1,
  ......// Author: Skip to save space
  output wire [31:0] y8,
  output wire done
);

// Please fill in the expression of twiddle factors
  localparam [31:0] Twiddle_Factor_W_8_0 = ...;
  ......// Author: Skip to save space
  localparam [31:0] Twiddle_Factor_W_8_3 = ...;

// Stage_1a
  wire stage_1a_done;
  wire stage_1a_enable = enable;
  wire [31:0] stage_1a_out1,......,stage_1a_out8;
  // Author: Skip to save space
	
  butterfly_computation butterfly_stage_1a_1( 
    .clk(clk), 
    .reset(reset), 
    .enable(stage_1a_enable), 
    .in1(x1), .in2(x5), 
    .out1(stage_1a_out1), .out2(stage_1a_out5), 
    .done() );
  ......// Author: skip 1a_2 and 1a_3 to save space
  butterfly_computation butterfly_stage_1a_4( 
    .clk(clk), 
    .reset(reset), 
    .enable(stage_1a_enable), 
    .in1(x4), .in2(x8), 
    .out1(stage_1a_out4), .out2(stage_1a_out8), 
    .done(stage_1a_done) );
\end{lstlisting}
\end{tcolorbox}
\end{figure}
\begin{figure}[htbp]
\centering
\begin{tcolorbox}[sharp corners, colback=white, colframe=black, left=2.5mm]
\begin{lstlisting}[firstnumber=40,style={verilog-style}]
// Stage_1b
  wire stage_1b_done;
  wire stage_1b_enable = stage_1a_done;
  wire [31:0] stage_1b_out1, ......, stage_1b_out8;
  // Author: Skip to save space
  
  assign stage_1b_out1 = stage_1a_out1;
  ......// Author: Skip to save space
  assign stage_1b_out4 = stage_1a_out4;
  
  complex_multiplication twiddle_stage_1b_1( 
    .clk(clk),
    .reset(reset),
    .enable(stage_1b_enable),
    .in1(stage_1a_out5),.in2(Twiddle_Factor_W_8_0),
    .out(stage_1b_out5),
    .done() );
  ......// Author: skip 1b_2 and 1b_3 to save space
  complex_multiplication twiddle_stage_1b_4( 
    .clk(clk),
    .reset(reset),
    .enable(stage_1b_enable),
    .in1(stage_1a_out8),.in2(Twiddle_Factor_W_8_3),
    .out(stage_1b_out8),
    .done(stage_1b_done) );
    
// Stage 2
  wire stage_2_done;
  assign stage_2_enable = stage_1b_done;
  wire [31:0] stage_2_out1, ......, stage_2_out8;
  // Author: Skip to save space
	
  fft_4_point fft_4_point_stage_2_1( 
    .clk(clk), 
    .reset(reset), 
    .enable(stage_2_enable),
    .x1(stage_1b_out1), .x2(stage_1b_out2), 
    .x3(stage_1b_out3), .x4(stage_1b_out4),
    .y1(stage_2_out1), .y2(stage_2_out2), 
    .y3(stage_2_out3), .y4(stage_2_out4), 
    .done() );
	    	
  fft_4_point fft_4_point_stage_2_2( 
    .clk(clk), 
    reset(reset), 
    .enable(stage_2_enable),
    .x1(stage_1b_out5), .x2(stage_1b_out6), 
    .x3(stage_1b_out7), .x4(stage_1b_out8),
    .y1(stage_2_out5), .y2(stage_2_out6), 
    .y3(stage_2_out7), .y4(stage_2_out8), 
    .done(stage_2_done) );

// Final output
  assign y1 = stage_2_out1;
  ......// Author: Skip to save space
  assign y8 = stage_2_out8;
  assign done = stage_2_done;	
endmodule
\end{lstlisting}
\end{tcolorbox}
\captionsetup{font={footnotesize}}
\caption{Example No.2: a functional eight-point FFT generated by ChatGPT. To conserve space, we omit portions of the code that are highly similar or repetitive (refer to our comments on Line 6, Line 9, Line 16, Line 23, Line 32, Line 44, Line 47, Line 57, Line 70, and Line 94 for more details).}\label{Fig_22}
\end{figure}

\begin{figure}[htbp]
  \centering
\begin{tcolorbox}[sharp corners, colback=white, colframe=black, left=2.5mm]
\begin{lstlisting}[style={verilog-style}]
module fft_64_point (
  input wire clk,
  input wire reset,
  input wire enable,
  input wire [31:0] x1,
  ......// Author: Skip to save space
  input wire [31:0] x64,
  output wire [31:0] y1,
  ......// Author: Skip to save space
  output wire [31:0] y64,
  output wire done
);

// Please fill in the expression of twiddle factors
  localparam [31:0] Twiddle_Factor_W_64_0 = ...;
  ......// Author: Skip to save space
  localparam [31:0] Twiddle_Factor_W_64_31 = ...;

// Stage 1a
  wire stage_1a_done;
  wire stage_1a_enable = enable;
  wire [31:0] stage_1a_out1,......,stage_1a_out64;
  // Author: Skip to save space
  butterfly_computation butterfly_stage_1a_1( 
    .clk(clk), 
    .reset(reset), 
    .enable(stage_1a_enable), 
    .in1(x1), .in2(x33), 
    .out1(stage_1a_out1), .out2(stage_1a_out33), 
    .done() );
  ......// Author: skip 1a_2 to 1a_31 to save space
\end{lstlisting}
\end{tcolorbox}
\end{figure}

\begin{figure}[htbp]
  \centering
\begin{tcolorbox}[sharp corners, colback=white, colframe=black, left=2.5mm]
\begin{lstlisting}[firstnumber=32, style={verilog-style}]	
  butterfly_computation butterfly_stage_1a_32( 
    .clk(clk), 
    .reset(reset), 
    .enable(stage_1a_enable), 
    .in1(x32), .in2(x64), 
    .out1(stage_1a_out32), .out2(stage_1a_out64), 
    .done(stage_1a_done) );
    
// Stage 1b
  wire stage_1b_done;
  wire stage_1b_enable = stage_1a_done;
  wire [31:0] stage_1b_out1,......,stage_1b_out64;

  assign stage_1b_out1 = stage_1a_out1;
  ......// Author: Skip to save space
  assign stage_1b_out32 = stage_1a_out32;

  complex_multiplication twiddle_stage_1b_1( 
    .clk(clk),
    .reset(reset),
    .enable(stage_1b_enable),
    .in1(stage_1a_out33),.in2(Twiddle_Factor_W_64_0),
    .out(stage_1b_out33),
    .done() );
  ......// Author: Skip to save space
  complex_multiplication twiddle_stage_1b_32( 
    .clk(clk),
    .reset(reset),
    .enable(stage_1b_enable),
    .in1(stage_1a_out64),.in2(Twiddle_Factor_W_64_31),
    .out(stage_1b_out64),
    .done(stage_1b_done) );
    
// Stage 2
  wire stage_2_done;
  wire stage_2_enable = stage_1b_done;
  wire [31:0] stage_2_out1,......,stage_2_out64;
  // Author: Skip to save space

  fft_32_point fft_32_point_stage_2_1( 
    .clk(clk), 
    .reset(reset), 
    .enable(stage_2_enable), 
    .x1(stage_1b_out1),
    ......// Author: Skip to save space
    .x32(stage_1b_out32), 
    .y1(stage_2_out1), 
    ......// Author: Skip to save space
    .y32(stage_2_out32), 
    .done() );
    
  fft_32_point fft_32_point_stage_2_2( 
    .clk(clk), 
    .reset(reset), 
    .enable(stage_2_enable), 
    .x1(stage_1b_out33), 
    ......// Author: Skip to save space
    .x32(stage_1b_out64), 
    .y1(stage_2_out33),  
    ......// Author: Skip to save space
    .y32(stage_2_out64), 
    .done(stage_2_done) );

// Final output
  assign y1 = stage_2_out1;
  ......// Author: Skip to save space
  assign y64 = stage_2_out64;
  assign done = stage_2_done;	
endmodule
\end{lstlisting}
\end{tcolorbox}
  \captionsetup{font={footnotesize}}
  \caption{A functional 64-point FFT generated by ChatGPT. To conserve space, we omit portions of the code that are highly similar or repetitive (refer to our comments on Line 6, Line 9, Line 16, Line 22, Line 31, Line 46, Line 56, Line 69, Line 76, Line 79, Line 87, Line 91, and Line 97 for more details).}\label{Fig_24}
\end{figure}

%% file: Source/Appendix_A6.tex
\begin{figure*}[htb]
  \centering
  \begin{tcolorbox}[sharp corners, colback=white, colframe=black, left=2.5mm]
\begin{lstlisting}[style={prompt-style}]
I am developing a 16-point FFT IP core on FPGA. In my design, the output of some butterfly computation need to do twiddling operation. In essence, the twiddling process
involves a complex multiplication. There should be two inputs for the complex multiplication. The first input is the output of the butterfly computation, which is
a 16-bit signal, with a 16-bit imag part and a 16-bit real part. The second input is the corresponding twiddle factor, which should also be a 16-bit signal. 
I want you to help me to calculate the twiddle factors for my 16-point FFT IP core.

For your reference, I can show you how I calculate the twiddle factors for an 8-point FFT. You should learn from the following steps.

Step 1: Since we are talking about an 8-point FFT, there should be 4 twiddle factors. Let us call them Twiddle_Factor_W_8_0, Twiddle_Factor_W_8_1, 
Twiddle_Factor_W_8_2, and Twiddle_Factor_W_8_3, respectively.

Step 2: In the twiddling operation, you should multiply four butterfly's output by exp(-j*2*pi*0/N), exp(-j*2*pi*1/N), 
exp(-j*2*pi*2/N), 
and exp(-j*2*pi*3/N), where N is 8 for the 8-point FFT. So the specific value of the above four numbers should be:
a)	number 1 = exp(-j*2*pi*0/N) = exp(-j*0*pi/8) = 1+0*i.
b)	number 2 = exp(-j*2*pi*1/N) = exp(-j*2*pi/8) = 0.7071 - 0.7071*i.
c)	number 3 = exp(-j*2*pi*2/N) = exp(-j*4*pi/8) = 0-1*i.
d)	number 4 = exp(-j*2*pi*3/N) = exp(-j6*pi/8) = -0.7071 - 0.7071*i.

Step 3: For the convenience of hardware signal processing, we need to scale number 1/2/3/4 by 2^15-1, which is 32767. 
So, now, we have
a)	number 1' = (1+0*i)* 32767 = 32767 + 0i.
b)	number 2' = (0.7071 - 0.7071*i)* 32767 = 23169.5457 - 23169.5457*i.
c)	number 3' = (0-1*i)* 32767= 0 - 32767*i.
d)	number 4' = (-0.7071 - 0.7071*i)* 32767 = -23169.5457 - 23169.5457*i.

Step 4: In hardware processing, we deal with integer. So, now, we have
a)	number 1'' = round(number 1') = 32767 + 0i.
b)	number 2'' = round(number 2') = 23170 - 23170*i.
c)	number 3'' = round(number 3') = 0 - 32767*i.
d)	number 4'' = round(number 4') = -23170 - 23170*i.

Step 5: we express the number 1''/2''/3''/4'' in the form of 32-bit complex number (with 16-bit imag part 
and 16-bit real-part):
a)	imag(Twiddle_Factor_W_8_0)=signed_dec_to_signed_bin( imag(number 1'') )=16b' 0000000000000000;
    real(Twiddle_Factor_W_8_0)=signed_dec_to_signed_bin( real(number 1'') )=16b' 0111111111111111.
Therefore, Twiddle_Factor_W_8_0={16'b0000000000000000,16'b0111111111111111};

b)	imag(Twiddle_Factor_W_8_1)=signed_dec_to_signed_bin( imag(number 2'') )=16b' 1010010101111110;
    real(Twiddle_Factor_W_8_1)=signed_dec_to_signed_bin( real(number 2'') )=16b' 0101101010000010.
Therefore, Twiddle_Factor_W_8_1={16'b1010010101111110,16'b0101101010000010};

c)	imag(Twiddle_Factor_W_8_2)=signed_dec_to_signed_bin( imag(number 3'') )=16b' 1000000000000001;
    real(Twiddle_Factor_W_8_2)=signed_dec_to_signed_bin( real(number 3'') )=16b' 0000000000000000.
Therefore, Twiddle_Factor_W_8_2={16'b0000000000000000,16'b1000000000000001};

d)	imag(Twiddle_Factor_W_8_3)=signed_dec_to_signed_bin( imag(number 4'') )=16b' 1010010101111110;
    real(Twiddle_Factor_W_8_3)=signed_dec_to_signed_bin( real(number 4'') )=16b' 1010010101111110.
Therefore,Twiddle_Factor_W_8_3={16'b1010010101111110,16'b1010010101111110};

Finally, my question is: Please generate the twiddle factors for my 16-point FFT module? There should be 8 values. You need to 
a)	give me the generation process of Twiddle_Factor_W_16_1
b)	list the final expressions of the 8 twiddle factors.
\end{lstlisting}
\end{tcolorbox}
  \captionsetup{font={footnotesize}}
  \caption{The CoT prompt we used to generate the 32-bit sequences for twiddle factors in a 16-point FFT.}\label{Fig_26}
\end{figure*}

%% file: Source/Appendix_B1.tex
\setcounter{figure}{0} 
This section presents how we conduct experiments to quantify the productivity gain of using LLMs to assist hardware development on FPGAs.

The benchmark experiment was conducted among four groups of students from the University of Electronic Science and Technology of China (UESTC), a well-known university with strong educational backgrounds in electronic engineering. Volunteers from Group One and Group Two are senior undergraduate students with course experiences in hardware design and Verilog HDL, while volunteers from Group Three and Group Four are senior postgraduate students from an FPGA\&ASIC-focused research lab. 

Among undergraduates, we have 22 volunteers in the non-GPT-assisted group (Group One) and 5 in the GPT-assisted group (Group Two). Among postgraduates, there are 9 volunteers in the non-GPT-assisted group (Group Three) and 4 in the GPT-assisted group (Group Four). The proportion of GPT users is lower among undergraduates than postgraduates because undergraduate GPT users are harder to find. This disparity can be attributed to the differing academic focuses: postgraduate students, often engaged in research, might resort to GPT more frequently than undergraduates, who primarily concentrate on coursework.

We designed three representative Verilog programming tasks for FPGA development (see subsection A below for detailed descriptions of these tasks) and asked the volunteers to complete the tasks. Volunteers from Group Two and Group Four were allowed to complete their tasks with the assistance of ChatGPT.\footnote{We gave volunteers from Group Two and Group Four a quick introduction to methods presented in this paper to ensure that they know how to utilize LLMs in FPGA development.} On the other hand, volunteers from Group One and Group Three, as comparisons, had to finish the tasks on their own. We graded each volunteer's answer in terms of time consumption and code quality (see subsection B for detailed grading standards). The results of the experiments are presented and analyzed in subsection C below.

\subsection{Tested Questions}
\begin{figure}[htbp]
  \centering
  \includegraphics[width=0.475\textwidth]{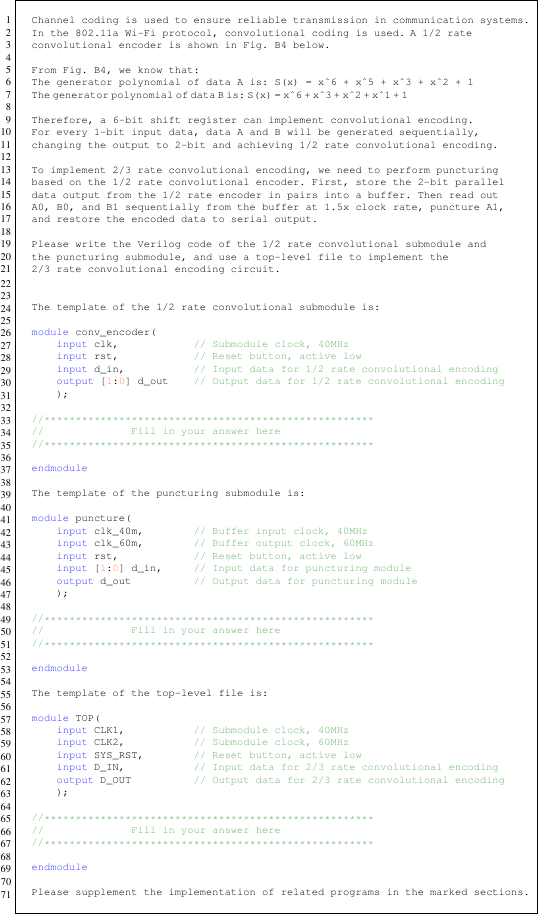}\\
  \captionsetup{font={footnotesize}}
  \caption{The requirement of Task Three.}\label{Task3}
\end{figure}

We designed the following three tasks for the experiments. Task 1 asks volunteers to write a testbench and validate a given Verilog module, while Task 2 and Task 3 focus on signal processing algorithms in wireless communication. Specifically, Task 2 asks volunteers to implement a simple second-order IIR filter, while Task 3 requires implementing a more complex convolution encoder with a 2/3 code rate. For undergraduate volunteers, we assigned Task 1 and Task 2; for graduate student volunteers, we assigned all three tasks. Fig. \ref{Task1}, \ref{Task2}, and \ref{Task3} present the detailed requirements of the three tasks.

\begin{figure}[htbp]
\centering
\begin{tcolorbox}[sharp corners, colback=white, colframe=black, left=2.5mm]
\begin{lstlisting}[style={verilog-style}]
We have implemented an LED scrolling light system controlled by buttons on an
FPGA development board. There is a reset button (i.e. button 1) plus 2 control
buttons (i.e. button 2, button 3). The system clock is 100MHz. There are 4 LED
lights that change every 0.2s.

The final displayed effect is: when no buttons are pressed, all LED lights are
on; when button 1 is pressed, the LED lights light up from right to left in 
serial; when button 2 is pressed, the LED lights light up from left to 
right in swrial.

The code of the button-controlled LED scrolling light is as follows:

module led(
    input sys_clk,          // 100MHz system clock, clock period 10ns  
    input sys_rst,          // Reset button, active low
    input [1:0] key,        // Button input signals   
    output reg [3:0] led    // LED output signals
    );
    
// Define a counter for clock division
reg [27:0] counter;
always@(posedge sys_clk or negedge sys_rst)
    begin
        if (!sys_rst)
            counter <= 28'd0;
        else if (counter < 28'd2000_0000)
            counter <= counter + 1'b1'
        else
            counter <= 28'd0;
    end
    
// Define a state variable to control LED
reg [1:0] led_state;
always@(posedge sys_clk or negedge sys_rst)
    begin
        if (!sys_rst)
            led_state <= 2'b00;
        else if (counter == 28'd1999_0000)
            led_state <= led_state + 1'b1'
        else
            led_state <= led_state;
    end

// Recognize button state, select scrolling light mode
always@(posedge sys_clk or negedge sys_rst)
    begin
        if (!sys_rst)
            led <= 4'b0000;
        else if (key[0] == 0) 
            // Press button 1, LED lights up from right to left in sequence
            case (led_state)      
                2'b00 : led <= 4'b0001;
                2'b01 : led <= 4'b0010;
                2'b10 : led <= 4'b0100;
                2'b11 : led <= 4'b1000;
                default : led <= 4'b0000;
            endcase
        else if (key[1] == 0)       
            // Press button 2, LED lights up from left to right in sequence
            case (led_state)
                2'b00 : led <= 4'b1000;
                2'b01 : led <= 4'b0100;
                2'b10 : led <= 4'b0010;
                2'b11 : led <= 4'b0001;
                default : led <= 4'b0000;
            endcase
        else
            // No buttons pressed, all LED lights on
            led <= 4'b1111;         
    end  
endmodule

The template of the testbench is as follows:

module tb_led();
//***********************************************
//            Fill in your answer here
//***********************************************
endmodule
\end{lstlisting}
\end{tcolorbox}
\captionsetup{font={footnotesize}}
\caption{The requirement of Task One}\label{Task1}
\end{figure}

\begin{figure}[htbp]
\centering
\centering
\begin{tcolorbox}[sharp corners, colback=white, colframe=black, left=2.5mm]
\begin{lstlisting}[style={verilog-style}]
IIR filters are commonly used in digital signal processing. Here we want an IIR 
filter with a cascaded structure. The difference equation of the filter is:

2048y(n) = 94x(n) + 140x(n-1) + 94x(n-2) + 1213y(n-1) - 268y(n-2)

Please realize the target IIR filter in Verilog. Here is the template:

module IIR_firstlayer (    
    input sys_clk,              // 2kHz system clock  
    input sys_rst,              // Reset button, active low
    input signed [11:0] Xin,    // Filter input data, input rate 2kHz 
    output signed [11:0] Yout   // Filter output data
    );
//***********************************************
//            Fill in your answer here
//***********************************************
endmodule
\end{lstlisting}
\end{tcolorbox}
\captionsetup{font={footnotesize}}
\caption{The requirement of Task Two.}\label{Task2}
\end{figure}

\begin{figure}[htbp]
  \centering
  \includegraphics[width=0.325\textwidth]{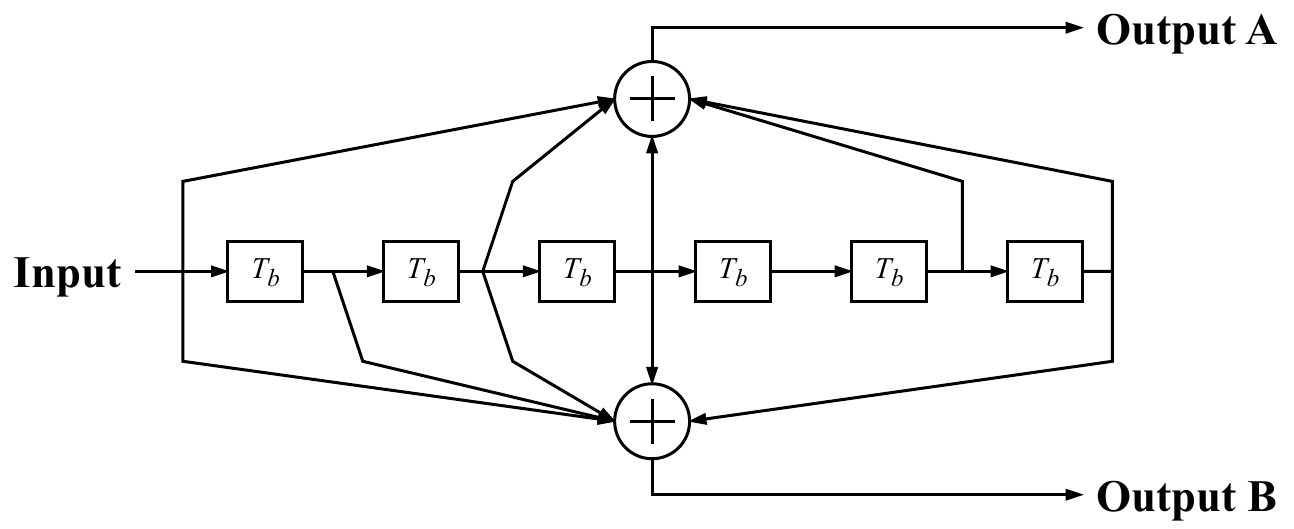}\\
  \captionsetup{font={footnotesize}}
  \caption{An illustration of the example in Fig. \ref{Task3}.}\label{Task3_Fig}
\end{figure}

%% file: Source/Appendix_B2.tex
\subsection{Grading Standard}
To comprehensively evaluate the quality of submitted Verilog code, we designed a scoring scheme to grade submissions based on four key criteria:
\begin{itemize}
\item \textbf{Reliability (30\%):} The code should pass testbench validation and return correct results. It should also avoid potential bugs like register overflow.
\item \textbf{Efficiency (30\%):} The code should minimize latency during calculations. Optimization strategies like ping-pong buffers are encouraged.
\item \textbf{Readability (20\%):} Well-commented and elegantly written code with descriptive variable names enhances understandability and efficiency. This allows code reviewers to easily grasp the logic.
\item \textbf{Maintainability (20\%):} The code should facilitate future changes and development. Modular, reusable functions are preferred.
\end{itemize}

Submissions are scored from 1 to 5 based on these metrics. To ensure fairness, all codes are blindly reviewed by the same Verilog expert.

%% file: Source/Appendix_B3.tex
\setcounter{table}{0} 
\begin{table*} [htbp]
\footnotesize 
\captionsetup{font={footnotesize}}
\caption{Benchmarking Results in Undergraduate and Postgraduate Groups}
\centering
\begin{tabular}{|c|c|c|c|c|c|c|c|}
\hline
& & Group 1 & Group 2 & Group 3 & Group 4 & LLM's Gain for undergraduate & LLM's Gain for postgraduate  \\
\hline
\multirow{2}*{Task 1} 
& Time Cons. & 53.39 min & 18.60 min & 23.44 min & 7.40 min & ${-65.16\%}$ & ${-68.44\%}$\\ 
\cline{2-8} 
& Quality Score & 4.31 / 5.00 & 4.76 / 5.00 & 4.62 / 5.00 & 4.78 / 5.00 & ${10.46\%}$ & ${3.31\%}$ \\ 
\hline
\multirow{2}*{Task 2} 
& Time Cons. & 90.64 min & 45.80 min & 55.34 min & 31.93 min & ${-49.47\%}$ & ${-42.32\%}$ \\ 
\cline{2-8} 
& Quality Score & 3.13 / 5.00 & 4.51 / 5.00 & 3.70 / 5.00 & 4.75 / 5.00 & ${44.22\%}$ & ${28.38\%}$ \\ 
\hline
\multirow{2}*{Task 3} 
& Time Cons. & / & / & 76.92 min & 34.13 min & / & ${-55.67\%}$ \\ 
\cline{2-8} 
& Quality Score & / & / & 3.39/5.00 & 4.45/5.00 & / & ${31.31\%}$ \\ 
\hline
\end{tabular}
\normalsize 
\end{table*}

\subsection{Results, Analysis, and Discussions}
We present our experimental results in Table B1. The performance of these volunteers across the three tasks is summarized as follows:

\begin{itemize}
\item For Task 1, the help of LLM greatly reduced the time consumption of undergraduate volunteers by 65.16\% and postgraduate volunteers by 68.44\%, demonstrating the impressive productivity gain of LLMs for testbench writing. Furthermore, LLMs also helped improve the quality of the testbench code, leading to a 10.46\% quality improvement for undergraduate volunteers and a 3.31\% improvement for postgraduate volunteers.
\item For Task 2, LLMs reduced the coding time of undergraduate volunteers by 49.47\% and postgraduate volunteers by 42.32\%, respectively. The quality score of the LLM-assisted group increased by 44.22\% for undergraduates and 28.38\% for postgraduates.
\item For Task 3, the most complex task specially designed for postgraduates, LLMs helped cut down the coding time by 55.67\% and increased the code quality by 31.31\%.
\end{itemize}

Apart from these straightforward observations, we also found some interesting phenomena in the data. First, we looked at the performance of Group 2 and Group 3 in Task 1 and Task 2. The data showed that the performance of volunteers in Group 2 was comparable to (or even slightly better than) that of volunteers in Group 3. \textit{This highlights the disruptive force of LLMs to bridge the gap between undergraduate and graduate students in task execution.}

We then examined the performance of Group 3 and Group 4 across the three tasks. A noteworthy insight emerges from the data. For Task 1, both Group 3 and Group 4 produced high-quality code, likely due to the simplicity of the task and postgraduate volunteers' superiority in such tasks. However, for the more complex Tasks 2 and 3, Group 4 demonstrated superior code quality. Upon consulting with the participants, this discrepancy was attributed to the typical engineering practice of prioritizing functional correctness with the expectation that subsequent reviews and edits would address any quality issues. On the other hand, the code produced by Group Two, which was created and refined with the assistance of LLMs, had already addressed many quality issues during the code generation stage itself. \textit{This indicates another LLM benefit: directly generating the desired code without the need for further polishing or editing}. Engineers in Group 3 need to invest additional time to review and edit their code manually.